\documentclass[12pt,a4paper]{article} 
\usepackage{graphicx}
\title{On the mathematical theory of news waves}

\author{Nikolay K. Vitanov $^{1,2}$*, Zlatinka I. Dimitrova$^{1}$, Kaloyan N. Vitanov$^1$}

\date{
$^{1}$ Institute of Mechanics, Bulgarian Academy of Sciences, Acad.
G. Bonchev Str., Bl. 4, 1113 Sofia, Bulgaria; vitanov@imbm.bas.bg\\
$^{2}$ Climate, Atmosphere and Water Research Institute, Bulgarian Academy of 
Sciences, Blvd. Tzarigradsko Chaussee 66, 1784 Sofia, Bulgaria\\
 }
\begin{document}
\maketitle

\begin{abstract}
We discuss the spread of a piece of news in a population. This is modeled by  SIR 
model of epidemic spread. The model  can be reduced to a nonlinear differential equation 
for the number of people affected by the   news of interest. The differential equation has an 
exponential nonlinearity and it  can be approximated by a sequence of nonlinear differential 
equations with polynomial  nonlinearities. Exact solutions to these equations can be obtained by the 
Simple Equations Method (SEsM). Some of these  exact solutions can be used to model a class of waves associated 
with the spread of the news in a population. The presence of exact 
solutions allow to study in detail the dependence of the amplitude and the time horizon of the news 
waves on the wave parameters such as the size of the population, initial number of spreaders of 
the piece of the news, transmission rate and recovery rate. This allows for recommendations about 
the change of wave parameters in order to achieve a large amplitude or appropriate time horizon of 
the news wave. We discuss 5 types of news waves on the basis of the values of the transmission rate
and recovery rate - the types A,B,C,D and E of news waves. In addition, we discuss the possibility
of building wavetrains by news waves. There are three possible kinds of wavetrains with respect
of the amplitude of the wave: increasing wavetrain, decreasing wavetrain, and mixed wavetrain.
The increasing wavetrain is especially interesting as it is connected to an increasing amplitude of
the news wave with respect to the amplitude of the previous wave of the wavetrain. It can find 
applications in advertising, propaganda, etc.
\end{abstract}

\section{Introduction}
The spread of news in a population is an important research problem \cite{news1,news2,news3} with
large significance for the practice. Examples are conspiracy theories 
\cite{news4,news4a,news4b,news4c}, echo chambers \cite{news5,news5a,news5b,news5c,news5d,news5e}, 
formation of mass opinion \cite{news6}, prejudice \cite{news7}, network propaganda \cite{news8}, 
exposure to ideologically diverse news \cite{news9}, etc. \cite{news10,news11}. Special attention is
given to the spread of misinformation and fake news 
\cite{news12,news13,news14,news15,news16,news17,news18,news19,
news20,news21,news22,news22a,news22b,news22c}. Let us also note the spread and misinformation and  fake news in the time of the COVID-19 pandemics 
\cite{news23,news24,news25,news26,news27,news28,news29,news30}.
\par 
Below we will use  results about the exact solutions connected to the SIR model of epidemics in
order to perform an analytical study of waves of news.  There exist various models for spreading of 
an epidemic in a population and a large amount of literature is devoted to this (for several examples 
see \cite{ep1} - \cite{ep10}). Epidemic models can also be applied for description of other
processes such as the spreading of ideas, for example (for overviews see \cite{vit16}, \cite{va12}). 
\par 
The text  is organized as follows. In Sect.2, we briefly discuss the use of the SIR model of 
epidemics spread to model the spread of a piece of news in a population as well as  
the reduction of the model equations to a chain of nonlinear differential equations. We present 
several exact solutions to the equations of this chain. The solutions are obtained by the Simple 
Equations Method (SEsM). The method is briefly described in Appendix A and the process of obtaining 
the solutions is illustrated in Appendix B. In Sect. 3, we use some of the obtained solutions
to derive analytical relationships for the three populations participating in the SIR model. In 
Sect. 4, we connect the obtained results to the spread of the news. Sect. 5 is devoted to a discussion of the possibilities for manipulation of the amplitude
and the time horizon of the news wave. We distinguish 5 types of news waves with respect to the
values of the transmission rate and recovery rate of the population where the wave spreads. Several concluding remarks are summarized in Sect. 6.
\section{The SIR Model of Epidemics as a Model for Spread of News}
The SIR model of epidemic spread can be reduced to a single nonlinear differential equation 
\cite{kmk,v23a}. Then, the obtained equation can be associated with a chain of nonlinear 
differential equations which contain polynomial nonlinearities. We will use this approach to obtain 
analytical solutions which can be connected to the spread of news waves.
\par 
The model of news waves is obtained as follows \cite{gior}. Let us consider a population of $N$ individuals.
The population is divided into 3 groups with respect to some piece of news. There is a subpopulation 
of the potential authors of spread of the piece of news - $S$. Then, there is a subpopulation of  
authors, active in posting piece of news - $I$. Finally, there is a subpopulation of authors, which 
became inactive in spreading of the piece of news after some period of 
activity in spreading that piece of news - $R$. The model equations 
for the time  change of the numbers of individuals from the above three 
subpopulations  are
\begin{eqnarray}\label{sir1}
\frac{dS}{dt} &=& - \frac{\tau}{N} S I \nonumber \\
\frac{dI}{dt} &=& \frac{\tau}{N} S I  - \rho I \nonumber \\
\frac{dR}{dt} &=& \rho I .
\end{eqnarray}
In (\ref{sir1}) $\tau$ is the transmission rate (quantitative characteristics of the transition
from the subpopulation of the potential spreaders to the subpopulation of active spreaders) and 
$\rho$ is the  recovery rate (quantitative characteristics of the transition from the subpopulation of active spreaders to the population of individuals who are not interested in the spread of the
piece of the news). We consider below the most simple case, where these rates are assumed to be constants. We are going to discuss analytical results for this simple case. These analytical results can serve as orientation for the numerical study of more complicated cases.
\par
From (\ref{sir1}) we obtain the relationship $N=S+I+R$. 
$N$ is the total population which is assumed to be constant. In several more words, we assume that 
the changes in the total population are negligible for the time of the studied phenomenon (the news 
wave).  We also stress the following. The system (\ref{sir1}) is written for
the spreaders of information. We can write the same system for people who hear some piece of news 
(hard news, soft news, fake news, etc.). In this case, $S$ will be the subpopulation of individuals 
who are susceptible to the news (who can hear the corresponding piece of news). $I$ will be the 
subpopulation of  individuals who have heard the piece of the news and are interested in spreading
that  piece of news. Finally, $R$ will be the subpopulation of individuals who are not interested 
anymore in the corresponding piece of news. In such a way, the model (\ref{sir1}) allows us to study
the process of spreading of the piece of news among the population of potentially interested people.
The news can be classified according to different characteristics. They can be hard news or soft 
news, true news or fake news. What will be of interest to us are the values of the transmission
rate $\tau$ and the recovery rate $\rho$. 
\par
Below, we will obtain analytical relationships for $R(t)$.
On the basis of these relationships, we can calculate the number
$I(t)$ from the SIR model.  This happens on the basis of 
the last equation of (\ref{sir1}):
\begin{equation}\label{sir3}
I = \frac{1}{\rho} \frac{dR}{dt}.
\end{equation}
On the basis of $I(t)$, we calculate the growth rate
\begin{equation}\label{sirx1}
\sigma(t) = \frac{1}{I} \frac{dI}{dt} .
\end{equation}
The growth rate $\sigma (t)$ can be written as $\sigma(t) = \rho(R_n - 1)$. 
Then
\begin{equation}\label{sirx3}
R_n(t) =  1+ \frac{\sigma(t)}{\rho}.
\end{equation} 
$R_n(t)$ is called  time varying effective reproduction number
for the spread of the news wave. There exists  a specific value: $R_n=1$. If $R_n<1$, then $
\sigma(t)<0$ and the  relative growth rate is negative. This means that $dI/dt$ is 
negative. In other words, the number of people, who post the piece of news (respectively the 
number of people who have heard the news and  are interested in that piece of news) will decrease 
and the  significance of the corresponding piece of news will decrease. If $R_n>1$, then $
\sigma(t)>0$ and the relative growth rate is positive. This means that $dI/dt$ is positive. In other 
words, the number of the people who post the piece of news (respectively the number of people who 
have heard the news and are interested in it) increases and the  significance of the 
corresponding piece of news increases. 
\par 
Another parameter of the news wave is its maximum $I_m$. This  maximum is achieved for some time 
$t_m$ after the initial moment. $t_m$ is called the time horizon of the news wave. The time horizon
of the news wave is an important characteristic and it will be mentioned many times below in the
text.
\par
Finally, the substitution of (\ref{sir3}) in the first equation of 
(\ref{sir1}) leads to the relationship
\begin{equation}\label{sir4}
S = S(0) \exp \left \{ - \frac{\tau}{\rho N} [R - R(0)] \right 
\}.
\end{equation}
Here $S(0)$ and $R(0)$ are the corresponding quantities at the time $t=0$. 
\par 
What remains, is to obtain an equation for $R(t)$. The substitution of $N=S+I+R$ and 
(\ref{sir4}) in the last equation of  (\ref{sir1}) leads to the differential equation for $R$
\begin{equation}\label{sir5}
\frac{dR}{dt} = \rho \left \{ N - R - S(0) \exp \left [ - 
\frac{\tau}{\rho N} (R - R(0)) \right ]  \right \}
\end{equation}
We assume $R(0)=0$ - no inactive news spreaders (respectively no individuals, which are uninterested in the piece of the news) at $t=0$. 
\par
The main assumption of the used approach is that the ratio $\frac{\tau R}{\rho N}$ is small enough.  
This can be realized, for an example when  $R$ is small enough in comparison to 
$N$.  In other words, the main assumption is that the news wave affects a relatively  small number of 
the  individuals of the population. For the case of such waves, $\exp \left [ - \frac{\tau}{\rho N} R  \right ]$ can  be represented as a Taylor series
$
\exp \left [ - \frac{\tau}{\rho N} R  \right ] = \sum 
\limits_{j=0}^M \frac{1}{j!}\left( - \frac{\tau}{\rho N} R \right)^j 
$.
The assumption here is that $\frac{e}{(M+1)!}\left( \frac{\tau R}{\rho N}\right)^{M+1}<<1$.
\par
$M$ has infinite value in the full Taylor series but we can 
truncate it at $M=2$, $M=3$,..., if  $- \frac{\tau}{\rho N} R$ is 
small enough. From (\ref{sir5}) we obtain
\begin{equation}\label{sir7}
\frac{dR}{dt} = \rho \left \{ N - R - S(0)  \sum \limits_{j=0}^M 
\frac{1}{j!}\left( - \frac{\tau}{\rho N} R \right)^j  \right \}, \ \ \ M=2,3,
\dots
\end{equation}
We set
\begin{equation}\label{sir8}
\alpha_0 = \rho[N-S(0)]; \ \ \ \alpha_1 = \frac{\tau S(0)}{N} - 
\rho; \ \ \ \alpha_j = -\frac{(-1)^j}{j!} 
\frac{\tau^j S(0)}{\rho^{j-1} N^j}, \ \ j=2,3,\dots 
\end{equation}
Then (\ref{sir7}) becomes
\begin{equation}\label{sir9}
\frac{dR}{dt} = \sum \limits_{j=0}^M \alpha_j R^j
\end{equation}
\par 
We assume that $\rho$ and $\tau$ are positive. Then $a_0$
will have a positive value. The value of $a_1$ can be positive or
negative. The values of $a_j$ is negative for even $j$ and positive
for odd $j$. We note that a similar reduction to a chain of
equations can also be made for the SEIR model of epidemic spread
\cite{v23b}.
\par
The chains of equations (\ref{sir7}) and (\ref{sir9}) are 
connected to the orders of approximation of (\ref{sir5}) which 
is the equation for the time evolution of the recovering
individuals for an epidemic wave within the scope of the SIR 
model. We can obtain exact solutions of these equations on
the basis of the Simple Equations Method (SEsM) (see the Appendix  \ref{app1}). The solutions which will be discussed here are listed
below. The detail about obtaining these solutions are given in Appendix
\ref{app2}. 
\par 
We will discuss the following solutions of the equations (\ref{sir9}).
For the case $M=2$ we discuss the solution
\begin{equation}\label{sir19}
R(t) =- \frac{\alpha_1}{2 \alpha_2} - \frac{\theta}{2 \alpha_2}
\tanh \left[ \frac{\theta(t+C)}{2} \right] + 
\frac{D}{\cosh^2\left[ \frac{\theta(t+C)}{2} \right] \left \{ E - 
\frac{2 \alpha_2 D}{\theta}
\tanh \left[ \frac{\theta(t+C)}{2} \right] \right \}} .
\end{equation}
We can write $G=D/E$ and this will reduce the number of the
constants of integration from 3 to 2 for the case $E \ne 0$.
In order to keep the case $E=0$ as a possibility, we are going to let
(\ref{sir19}) in the form without introduction of $G$. We note that
when $D=0$, one obtains the specific solution known since \cite{kmk}.
\par
For the cases $M>2$ we will discuss below solutions obtained for $L=1$.
For the case $M=3$, $L=1$ we discuss the solution  
\begin{eqnarray}\label{sir24}
R (t) = - \frac{\alpha_2}{3 \alpha_3} + \beta_1 \left \{ \frac{3 \alpha_1 \alpha_3 - \alpha_2^2}{  - 3 \alpha_3^2 \beta_1^2 + C 
(3 \alpha_1 \alpha_3 - \alpha_2^2)  \exp \{-2\frac{3 \alpha_1 \alpha_3 - \alpha_2^2}{3 \alpha_3}t \}} \right \}^{ \frac{1}{2}}
\end{eqnarray}
\par
We can write additional solutions for larger values of $M$.
For example, let us consider the case $M=4$, $L=1$. The solution is 
\begin{eqnarray}\label{sir33}
R (t) = -\frac{\alpha_3}{4 \alpha_4} + \beta_1 \left \{ \frac{  \alpha_3^4 - 256 \alpha_0 \alpha_4^3}{  64 \beta_1^3 \alpha_3  \alpha_4^3  + C( 
 \alpha_3^4 - 256 \alpha_0 \alpha_4^3)  \exp \{3   \frac{\alpha_3^4 - 256 \alpha_0 \alpha_4^3}{64 \alpha_3 \alpha_4^2}t \}} \right \}^{ \frac{1}{3}}.
\end{eqnarray}
\par 
Next, we consider the case $M=5$, $L=1$. The solution is
\begin{eqnarray}\label{sir38}
R= - \frac{\alpha_4}{5 \alpha_5} + \beta_1 \left \{ \frac{-\alpha_4^5+3125 \alpha_0 \alpha_5^4}{  - 625  \beta_1^4 \alpha_4 \alpha_3^4 + C (-\alpha_4^5+3125 \alpha_0 \alpha_5^4)  \exp \{-4\frac{-\alpha_4^5+3125 \alpha_0 \alpha_5^4}{625 \alpha_4 \alpha_5^3}t \}} \right \}^{ \frac{1}{4}}
\end{eqnarray}
\par 
The obtaining of exact solutions of the chain of equations can be 
continued. Below we focus on the properties of the news waves
described by the solutions (\ref{sir19}) and (\ref{sir24}). The solutions (\ref{sir33}) and
(\ref{sir38}) are also possible solutions for the specific cases of the studied chain of equations.
The application of these solutions to the situations modelled by the SIR model is limited
because of the relative large number of relationships among the parameters $a_i$ (see Appendix B).
Because of this we will not discuss them below.
\section{Discussion of the obtained exact analytical solutions of  the studied chain of equations}
\par 
Above, we have presented analytical relationships for the quantity $R(t)$.
This allows us to calculate the time evolution of the active in
the posting (interesting in the piece of news)
persons $I$ on the basis of (\ref{sir3}). Then we can calculate 
the relative growth rate from (\ref{sirx1}) and $R_n(t)$ from
(\ref{sirx3}). Our basic approximation for the 
reducing the SIR model to a chain of equations was $
\frac{e}{(M+1)!}\left(\frac{\tau R}{\rho N}\right)^{M+1} << 1$.  This means that the news wave has to affect a 
relatively small number of the entire population. If this is not the case, 
we have to solve the SIR model numerically. 
\par 
We have analytical relationships for several news waves. 
Thus, we can calculate their characteristics by means of the 
relationships, mentioned above.  For the calculation of $S$ we use the 
approximate relationship which occurs from (\ref{sir4})
\begin{equation}\label{sirx3a}
S(t) = S(0) \left [ 1 - \frac{\tau R}{\rho N} \right ]
\end{equation}
\par
We start from the specific solution (\ref{sir19}). From the requirement 
$R(0)=0$, we obtain for the constant of integration $C$:
\begin{eqnarray}\label{sir11y}
C= \frac{2}{\theta} {\rm atanh} \Bigg[  \frac{\theta (\alpha_1 E - 2 \alpha_2 D)}{2 \alpha_1 \alpha_2 D - \theta^2 E} \Bigg].
\end{eqnarray}
The solution (\ref{sir19}) becomes
\begin{eqnarray}\label{sirx12}
R(t) = - \frac{\alpha_1}{2 \alpha_2} - \frac{\theta}{2 \alpha_2}
\tanh \left[ \frac{\theta \Bigg \{t+\frac{2}{\theta} {\rm atanh} \Bigg[  \frac{\theta (\alpha_1 E - 2 \alpha_2 D)}{2 \alpha_1 \alpha_2 D - \theta^2 E} \Bigg]\Bigg \}}{2} \right] + \nonumber \\
\frac{D}{\cosh^2\left[ \frac{\theta \Bigg \{ t+\frac{2}{\theta} {\rm atanh} \Bigg[  \frac{\theta (\alpha_1 E - 2 \alpha_2 D)}{2 \alpha_1 \alpha_2 D - \theta^2 E} \Bigg] \Bigg \}}{2} \right] \left \{ E - 
\frac{2 \alpha_2 D}{\theta}
\tanh \left[ \frac{\theta \Bigg \{ t+\frac{2}{\theta} {\rm atanh} \Bigg[  \frac{\theta (\alpha_1 E - 2 \alpha_2 D)}{2 \alpha_1 \alpha_2 D - \theta^2 E} \Bigg] \Bigg \}}{2} \right] \right \}}
\end{eqnarray}
 (\ref{sirx12}) allows us to 
calculate the other quantities connected to this solution as 
follows
\begin{eqnarray}\label{sirx13}
I = \frac{1}{\rho}\frac{dR}{dt} = \frac{1}{\rho} \Bigg \{ \frac{\theta^2}{4 \alpha_2} \Bigg \{ 1- \tanh^2  \left[ \frac{\theta \Bigg \{t+\frac{2}{\theta} {\rm atanh} \Bigg[  \frac{\theta (\alpha_1 E - 2 \alpha_2 D)}{2 \alpha_1 \alpha_2 D - \theta^2 E} \Bigg]\Bigg \}}{2} \right]  \Bigg \} -
\nonumber \\ 
\frac{D \theta \tanh \left[ \frac{\theta \Bigg \{t+\frac{2}{\theta} {\rm atanh} \Bigg[  \frac{\theta (\alpha_1 E - 2 \alpha_2 D)}{2 \alpha_1 \alpha_2 D - \theta^2 E} \Bigg]\Bigg \}}{2} \right] \Bigg \{ 1 - \tanh^2 \left[ \frac{\theta \Bigg \{t+\frac{2}{\theta} {\rm atanh} \Bigg[  \frac{\theta (\alpha_1 E - 2 \alpha_2 D)}{2 \alpha_1 \alpha_2 D - \theta^2 E} \Bigg]\Bigg \}}{2} \right] \Bigg \}}{E - \frac{2 \alpha_2 D \tanh \left[ \frac{\theta \Bigg \{t+\frac{2}{\theta} {\rm atanh} \Bigg[  \frac{\theta (\alpha_1 E - 2 \alpha_2 D)}{2 \alpha_1 \alpha_2 D - \theta^2 E} \Bigg]\Bigg \}}{2} \right] }{\theta}} + \nonumber \\
 D^2 \alpha_2 \frac{\Bigg \{ 1- \tanh^2 \left[ \frac{\theta \Bigg \{t+\frac{2}{\theta} {\rm atanh} \Bigg[  \frac{\theta (\alpha_1 E - 2 \alpha_2 D)}{2 \alpha_1 \alpha_2 D - \theta^2 E} \Bigg]\Bigg \}}{2} \right]  \Bigg \}^2}{\Bigg \{ E-\frac{2 \alpha_2 D}{\theta} \tanh \left[ \frac{\theta \Bigg \{t+\frac{2}{\theta} {\rm atanh} \Bigg[  \frac{\theta (\alpha_1 E - 2 \alpha_2 D)}{2 \alpha_1 \alpha_2 D - \theta^2 E} \Bigg]\Bigg \}}{2} \right] \Bigg \}^2} \Bigg \}
\end{eqnarray}
In addition,
\begin{eqnarray}\label{sirx14}
S(t) = S(0) \Bigg \{ 1 - \frac{\tau}{\rho N } \Bigg\{ - \frac{\alpha_1}{2 \alpha_2} - \frac{\theta}{2 \alpha_2}
\tanh \left[ \frac{\theta \Bigg \{t+\frac{2}{\theta} {\rm atanh} \Bigg[  \frac{\theta (\alpha_1 E - 2 \alpha_2 D)}{2 \alpha_1 \alpha_2 D - \theta^2 E} \Bigg]\Bigg \}}{2} \right] + \nonumber \\
\frac{D}{\cosh^2\left[ \frac{\theta \Bigg \{ t+\frac{2}{\theta} {\rm atanh} \Bigg[  \frac{\theta (\alpha_1 E - 2 \alpha_2 D)}{2 \alpha_1 \alpha_2 D - \theta^2 E} \Bigg] \Bigg \}}{2} \right] \left \{ E - 
\frac{2 \alpha_2 D}{\theta}
\tanh \left[ \frac{\theta \Bigg \{ t+\frac{2}{\theta} {\rm atanh} \Bigg[  \frac{\theta (\alpha_1 E - 2 \alpha_2 D)}{2 \alpha_1 \alpha_2 D - \theta^2 E} \Bigg] \Bigg \}}{2} \right] \right \}}  \Bigg \}
\Bigg \} \nonumber \\
\end{eqnarray}
$\sigma(t)$ and $R_n(t)$ can be calculated from (\ref{sirx1}) and (\ref{sirx3}).
\par
The above results are valid if $\frac{e}{3!}\left( \frac{\tau R}{\rho N} 
\right)^3 \approx \left( \frac{\tau R}{\rho N} 
\right)^3<< 1$.
This means that
\begin{eqnarray}\label{check}
\frac{\tau^2}{\rho^2 N^2} \Bigg \{ - \frac{\alpha_1}{2 \alpha_2} - \frac{\theta}{2 \alpha_2}
\tanh \left[ \frac{\theta \Bigg \{t+\frac{2}{\theta} {\rm atanh} \Bigg[  \frac{\theta (\alpha_1 E - 2 \alpha_2 D)}{2 \alpha_1 \alpha_2 D - \theta^2 E} \Bigg]\Bigg \}}{2} \right] + \nonumber \\
\frac{D}{\cosh^2\left[ \frac{\theta \Bigg \{ t+\frac{2}{\theta} {\rm atanh} \Bigg[  \frac{\theta (\alpha_1 E - 2 \alpha_2 D)}{2 \alpha_1 \alpha_2 D - \theta^2 E} \Bigg] \Bigg \}}{2} \right] \left \{ E - 
\frac{2 \alpha_2 D}{\theta}
\tanh \left[ \frac{\theta \Bigg \{ t+\frac{2}{\theta} {\rm atanh} \Bigg[  \frac{\theta (\alpha_1 E - 2 \alpha_2 D)}{2 \alpha_1 \alpha_2 D - \theta^2 E} \Bigg] \Bigg \}}{2} \right] \right \}} \Bigg \}^3 <<1
\end{eqnarray}
For $t=0$, $R(0)=0$ and $\left( \frac{\tau R}{\rho N} 
\right)^3 << 1$ is satisfied. For $t \to \infty$, $R \approx - \frac{\alpha_1}{2 \alpha_2} - \frac{\theta}{2 \alpha_2}$, and 
$\left( \frac{\tau R}{\rho N}  \right)^3 << 1$ can be satisfied.
\par
Next, we discuss the solution   (\ref{sir24})
In this case, we have one relationship among $\alpha_0$, $\alpha_1$, $\alpha_2$ and $\alpha_3$. This leads to the following relationship for $\rho$
\begin{equation}\label{bern1}
\rho = \tau \left[1 -  \frac{S(0)}{3N} \right]
\end{equation}
For the constant of integration $C$ we obtain
\begin{equation}\label{bern2}
C=54 \beta_1^2 \frac{N-S(0)}{54 N^3 - 5 S(0)^3 + 36 N S(0)^2 - 81 S(0)N^2}
\end{equation}
This leads to the following result for $R(t)$
\begin{eqnarray}\label{bern3}
R(t)=\Bigg[ N - \frac{S(0)}{3} \Bigg] \Bigg \{ 1 - \Bigg[ \frac{6N - 5 S(0)}{S(0)+6[N-S(0)] \exp \Bigg (\frac{[6N -5S(0)] \tau t}{3N} \Bigg) } \Bigg ]^{1/2}\Bigg \}.
\end{eqnarray}
Then,
\begin{equation}\label{bern4}
I(t) =  \frac{[N-S(0)][6N-5S(0)]^{3/2}\exp \Bigg (\frac{[6N -5S(0)] \tau t}{3N} \Bigg)}{\Bigg[ S(0)+6[N-S(0)] \exp \Bigg (\frac{[6N -5S(0)] \tau t}{3N} \Bigg)\Bigg ]^{3/2}}
\end{equation}
\begin{equation}\label{bern5}
S \approx S(0)\Bigg\{ 1- \frac{\tau}{\rho N} \Bigg[ N - \frac{S(0)}{3} \Bigg] \Bigg \{ 1 - \Bigg[ \frac{6N - 5 S(0)}{S(0)+6[N-S(0)] \exp \Bigg (\frac{[6N -5S(0)] \tau t}{3N} \Bigg) } \Bigg ]^{1/2}\Bigg \} \Bigg \}
\end{equation}
On the bass of the obtained relationships, we can easily calculate
$\sigma(t)$ and $R_n(t)$ by use of (\ref{sirx1}) and (\ref{sirx3}).
\par 
Next, we have to check if the solution(\ref{bern3}) satisfies
the condition $\frac{e}{4!} \left(\frac{\tau R}{\rho N}\right)^5 << 1$.
At $t=0$, the condition is satisfied as $R(0)=0$. At $t \to \infty$, we have $R \approx \Bigg[ N - 
\frac{S(0)}{3} \Bigg]$. The condition is equal to $\frac{e}{4!}$, which is about $0.113$ and we can 
assume that this is small enough in comparison to $1$.  
\section{Characteristics of the news waves based on the solution  (\ref{sirx12})}
We proceed as follows. First, we discuss the case $D=0$. Then
we discuss the  corrections connected to the case $D\ne 0$.
\subsection{$D=0$}
In this case,  (\ref{sirx12}) becomes
\begin{eqnarray}\label{nw1}
R(t) = - \frac{\alpha_1}{2 \alpha_2} - \frac{\theta}{2 \alpha_2}
\tanh \left \{  \frac{\theta}{2} \Bigg [t+\frac{2}{\theta} {\rm atanh} \Bigg(  -\frac{\alpha_1}{\theta} \right) \Bigg ] \Bigg \}
\end{eqnarray}
In order to use this solution for the study of news waves we have to substitute
the relationships  for the coefficients $\alpha_i$ and $\theta$ in it.
For $\theta$, we obtain 
\begin{equation}\label{theta}
\theta = \Bigg[ \Bigg(\frac{\tau S(0)}{N} - \rho \Bigg)^2 +2 \frac{\tau^2 S(0)[N-S(0)]}{N^2}\Bigg]^{1/2}.
\end{equation}
The result for $R(t)$ is 
\begin{eqnarray}\label{nw2}
R(t) = \frac{\rho N^2}{\tau^2 S(0)}\Bigg \{  \frac{\tau S(0)}{N} - \rho   + \Bigg[ \Bigg(\frac{\tau S(0)}{N} - \rho \Bigg)^2 +2 \frac{\tau^2 S(0)[N-S(0)]}{N^2}\Bigg]^{1/2} \tanh \Bigg \{ t+ \nonumber \\
\frac{2}{\Bigg[ \Bigg(\frac{\tau S(0)}{N} - \rho \Bigg)^2 +2 \frac{\tau^2 S(0)[N-S(0)]}{N^2}\Bigg]^{1/2}} {\rm atanh} \Bigg \{\frac{\rho - \frac{\tau S(0)}{N} }{\Bigg[ \Bigg(\frac{\tau S(0)}{N} - \rho \Bigg)^2 +2 \frac{\tau^2 S(0)[N-S(0)]}{N^2}\Bigg]^{1/2}} \Bigg \} \Bigg \} \Bigg \}
\end{eqnarray}
Next, $I$ can be obtained by substitution of $\alpha_0$, $\alpha_1$, $\alpha_2$ and $\theta$ in (\ref{sirx13}).
And finally, we have
\begin{eqnarray}\label{nw4}
S \approx S(0) \Bigg \{ 1- \frac{N}{\tau S(0)} \Bigg \{  \frac{\tau S(0)}{N} - \rho   + \Bigg[ \Bigg(\frac{\tau S(0)}{N} - \rho \Bigg)^2 +2 \frac{\tau^2 S(0)[N-S(0)]}{N^2}\Bigg]^{1/2} \tanh \Bigg \{ t+ \nonumber \\
\frac{2}{\Bigg[ \Bigg(\frac{\tau S(0)}{N} - \rho \Bigg)^2 +2 \frac{\tau^2 S(0)[N-S(0)]}{N^2}\Bigg]^{1/2}} {\rm atanh} \Bigg \{\frac{\rho - \frac{\tau S(0)}{N} }{\Bigg[ \Bigg(\frac{\tau S(0)}{N} - \rho \Bigg)^2 +2 \frac{\tau^2 S(0)[N-S(0)]}{N^2}\Bigg]^{1/2}} \Bigg \} \Bigg \} \Bigg \} \Bigg \}
\end{eqnarray}

\begin{figure}[htb!]
\centering
\includegraphics[angle=-90,scale=0.4]{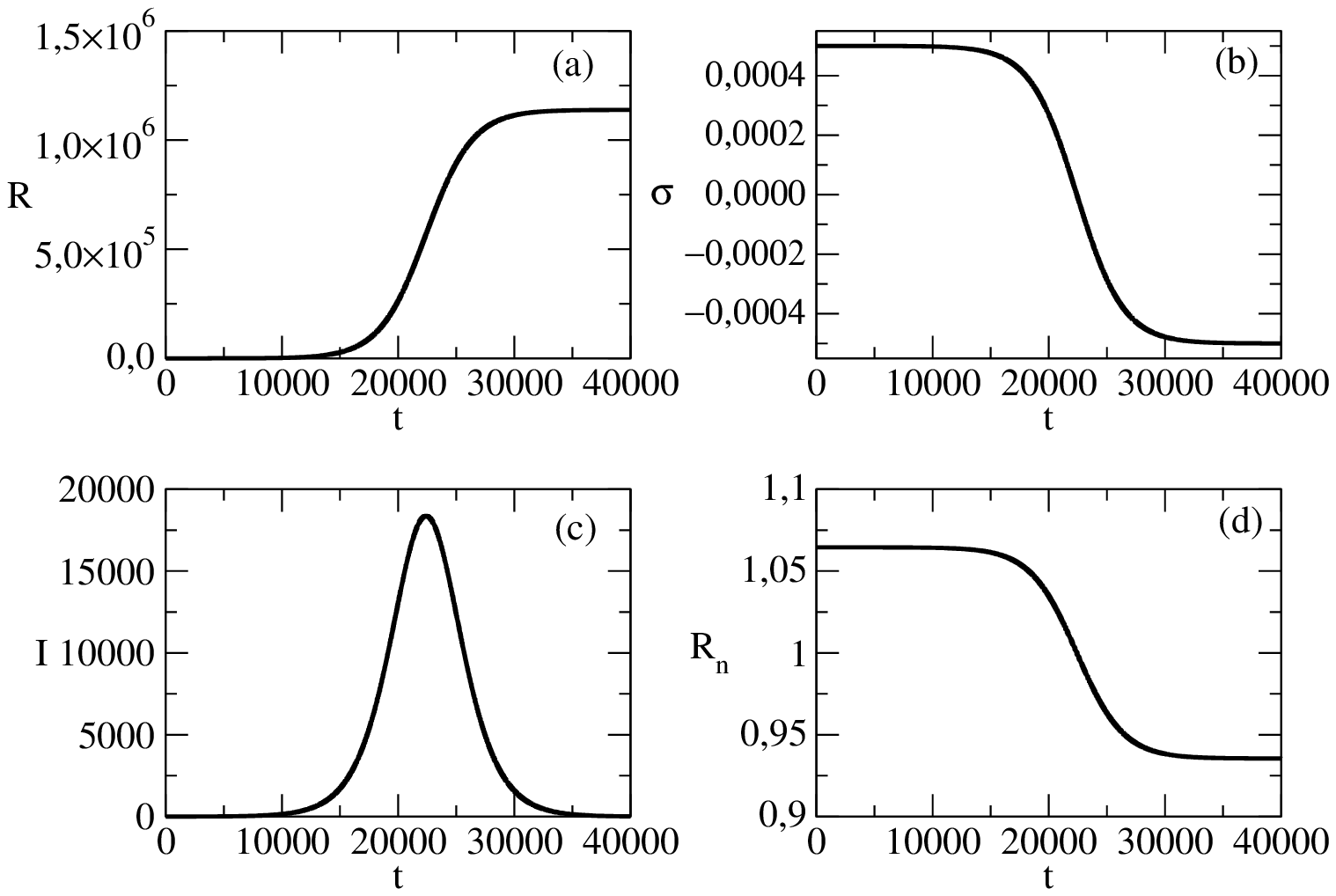}
\caption{The basic solution for all figures below. $N=10^7$, $S(0)=9 999 999$, $\tau=0.00825 $, $\rho=0.00775$. Figure (a): $R(t)$. Figure (b): $\sigma(t)$. Figure (c): $I(t)$. Figure (d): $R_n(t)$.}
\end{figure}
Figure 1 shows the basic solution for all of the following figures where we will show the influence of the changes of parameters on this basic solution. Figure (a) 
shows the number of individuals affected by the news wave. The number of individuals who are active 
in posting the piece of news is shown in Fig. (c). Figure (b) shows the 
growth rate $\sigma$ for the basic solution. Figure $R$ shows the effective
reproduction rate for the spread of the wave connected to the piece of the news.
\par 
\begin{figure}[htb!]
\centering
\includegraphics[angle=-90,scale=0.4]{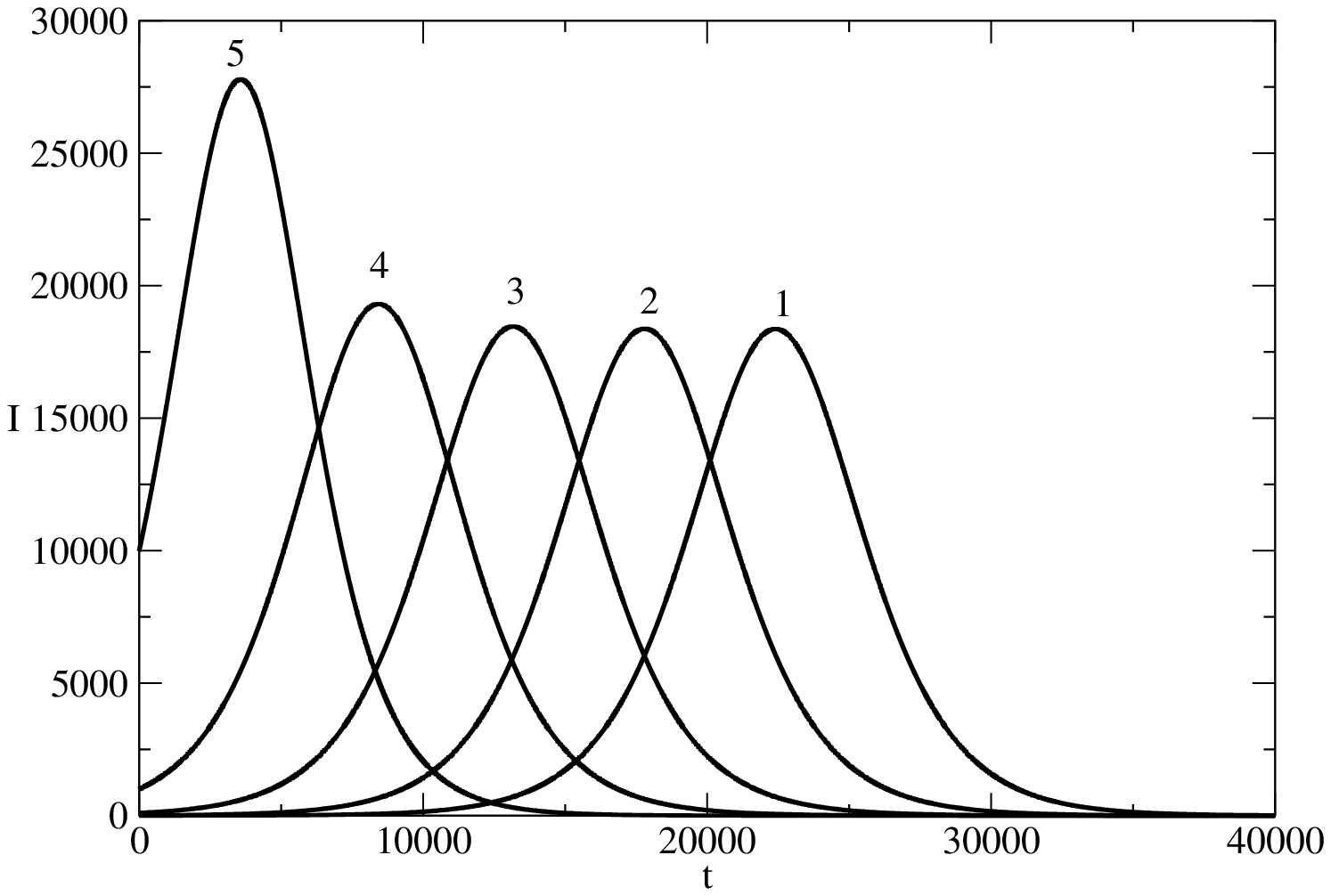}
\caption{Influence of $S(0)$ on the news wave. The profile of $I$ for the basic solution  $N=10^7$, 
$S(0)=9 999 999$, $\tau=0.00825 $, $\rho=0.00775$ is
denoted by 1. In other curves we change only the values of $S(0)$. 
Curve 2: $S(0)=9 999 990$. Curve 3: $S(0)=9 999 900$. Curve 4: $S(0) = 9 999 000$. Curve 5: $S(0)=9 990 000$.}
\end{figure}
Figure 2 shows the influence of the parameter $S(0)$ on the news wave.
As $R(0)=0$, then $I(0)=N-S(0)$ and actually , the figure shows the 
influence of the initial number of individuals who start to spread the piece 
of news on the evolution of the number of individuals who spread the
piece of news at the time of the existence of the news wave. Curve $1$
is for the case of the basic solution, which is for $I(0)=1$ individual
who starts to spread the piece of the news. Curve $2$ is for $I(0)=10$
individuals who start to spread the piece of news. We see that with an increase
of $I(0)$ there is a tendency to increase in the amplitude of the news wave 
and the maximum of the wave occurs earlier (the time horizon of the wave is shorter). 
This means that by manipulation of $I(0)$ one can have a news wave of certain amplitude at
a selected moment in the time. 
\par 
\begin{figure}[htb!]
\centering
\includegraphics[angle=-90,scale=0.4]{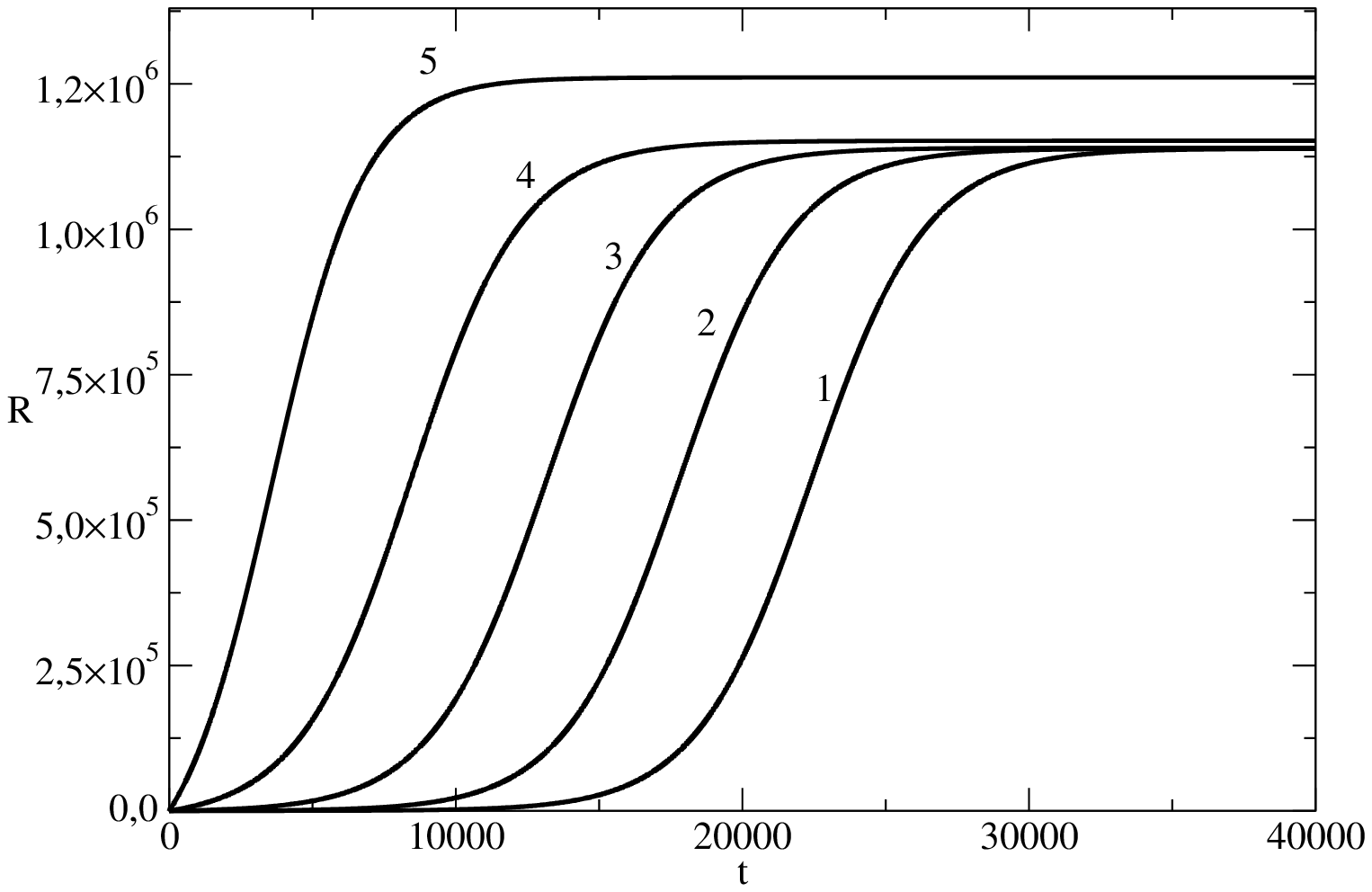}
\caption{Influence of $S(0)$ on the number $R$ of the individuals
affected by the news wave. $R$ for the basic solution  $N=10^7$, $S(0)=9 999 999$, $\tau=0.00825 $, $\rho=0.00775$ is
denoted by 1. In other curves we change only the values of $S(0)$. 
Curve 2: $S(0)=9 999 990$. Curve 3: $S(0)=9 999 900$. Curve 4: $S(0) = 9 999 000$. Curve 5: $S(0)=9 990 000$.}
\end{figure}
\par 
The influence of the initial number of individuals which start to spread the piece of news on the 
number of individuals affected by the news wave is shown in Fig. 3. It can be seen that the number 
of affected individuals increases by decreasing $S(0)$ (i.e., with an increasing $I(0)$).
\par 
\begin{figure}[htb!]
\centering
\includegraphics[angle=-90,scale=0.4]{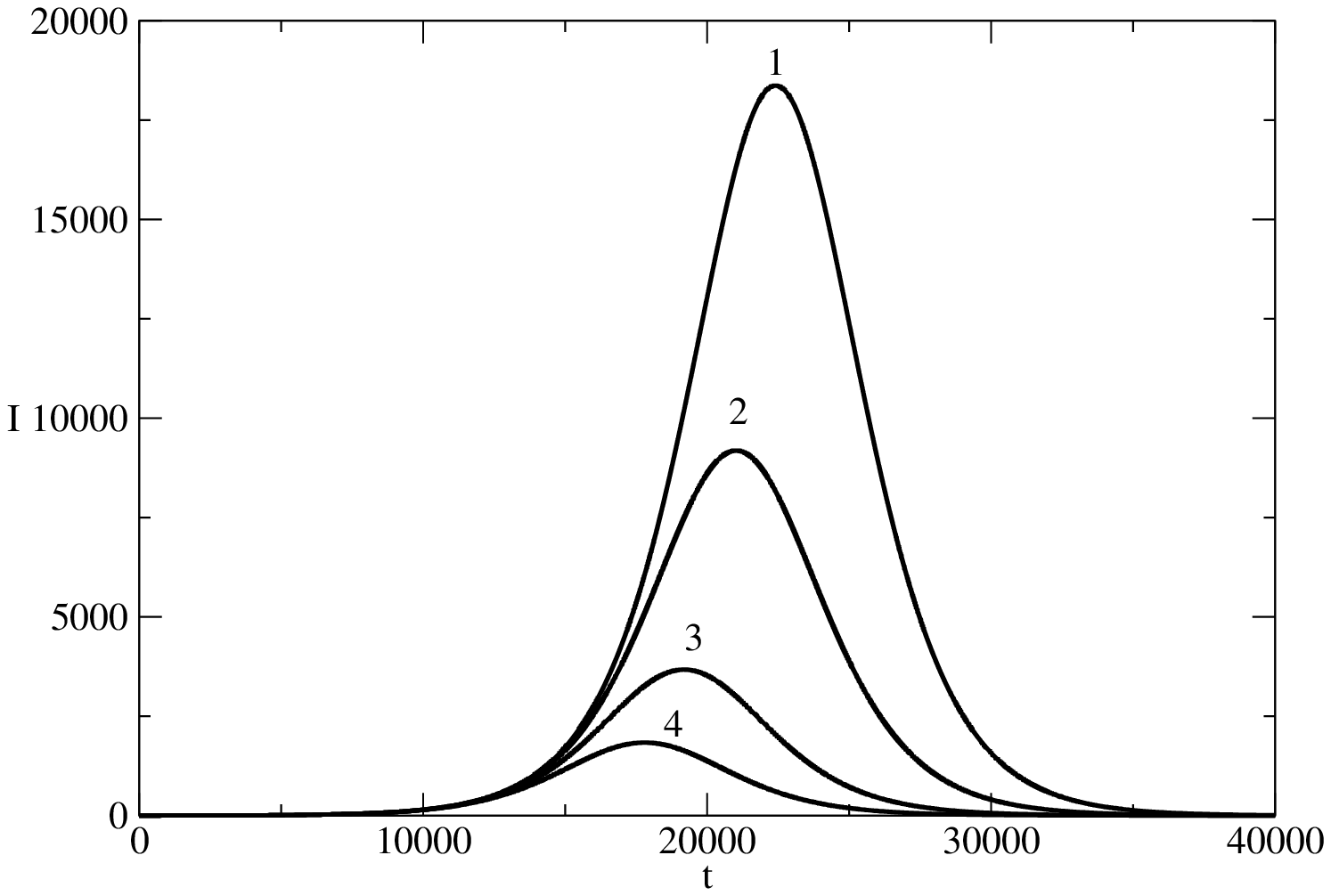}
\caption{Influence of the population $N$ on the news wave when
$S(0)=N-1$ (i.e., $I(0)=1$) on the number $I$ of the individuals
who spread the piece of the news.   $I$ for the the basic solution  $N=10^7$, $S(0)=9 999 999$, $\tau=0.00825 $, $\rho=0.00775$ is
denoted by 1. In other curves we change only the values of $N$ and $S(0)=N-1$. 
Curve 2: $N=5 000 000$. Curve 3: $N=2 500 000$. Curve 4: $N = 1 000 000$. }
\end{figure}
\par 
Figure 4 shows the influence of the population $N$ on the size of the news wave in the case when 
$I(0)=1$. In other words, when a single individual starts to spread the piece of news at the 
beginning of the wave. We observe two effects. First of all, a larger population leads to a larger 
wave amplitude. The second effect is quite interesting: the larger population leads to  slower wave 
and the peak of the wave moves to larger time (the time horizon of the wave is larger).
This means, that for the same wave parameters, in larger countries, the 
news waves peak is larger and the news wave exists for more time in comparison to the
corresponding news wave in a country with a smaller population.
\par 
\begin{figure}[htb!]
\centering
\includegraphics[angle=-90,scale=0.4]{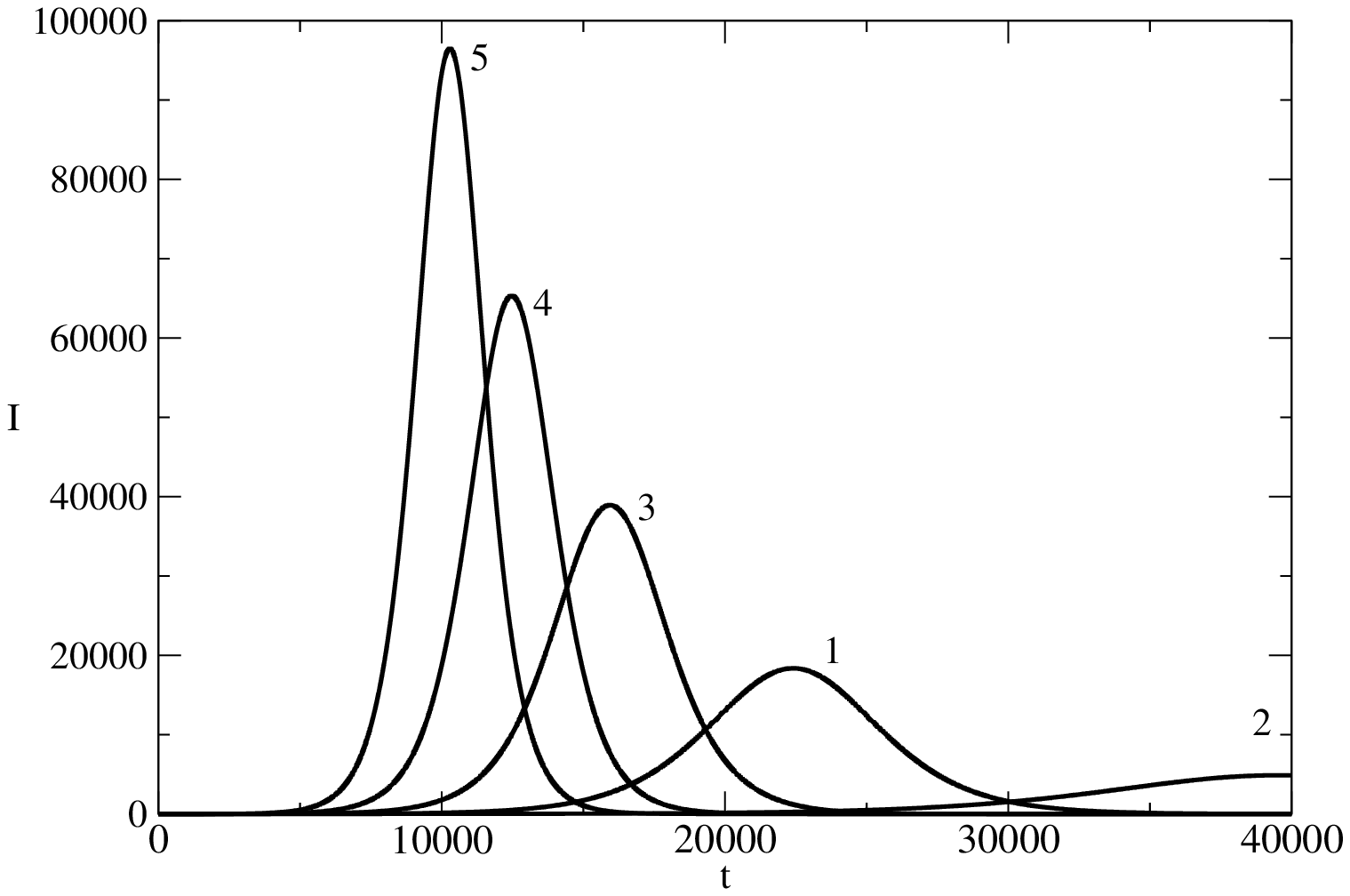}
\caption{Influence of the transmission rate $\tau$ on the news wave.   $I$ for the the basic solution  $N=10^7$, $S(0)=9 999 999$, $\tau=0.00825 $, $\rho=0.00775$ is
denoted by 1. In other curves we change only the values of $\tau$. 
Curve 2: $\tau=0.008$. Curve 3: $\tau=0.0085$. Curve 4: $\tau=0.0875$.
Curve 5: $\tau=0.009$. }
\end{figure}
\par 
Figure 5 shows the influence of the transmission rate $\tau$ on the news wave. Larger transmission
rate means that the piece of news spread easily in the corresponding population.
The basic solution is marked by 1 on the figure. We see that the decrease in
the value of the transmission rate leads to a news wave of smaller amplitude. In addition, the news 
wave develops slowly in time. The increase of the transmission rate leads to news waves of larger 
amplitude. The time horizon of the  wave becomes smaller.
\par 
\begin{figure}[htb!]
\centering
\includegraphics[angle=-90,scale=0.4]{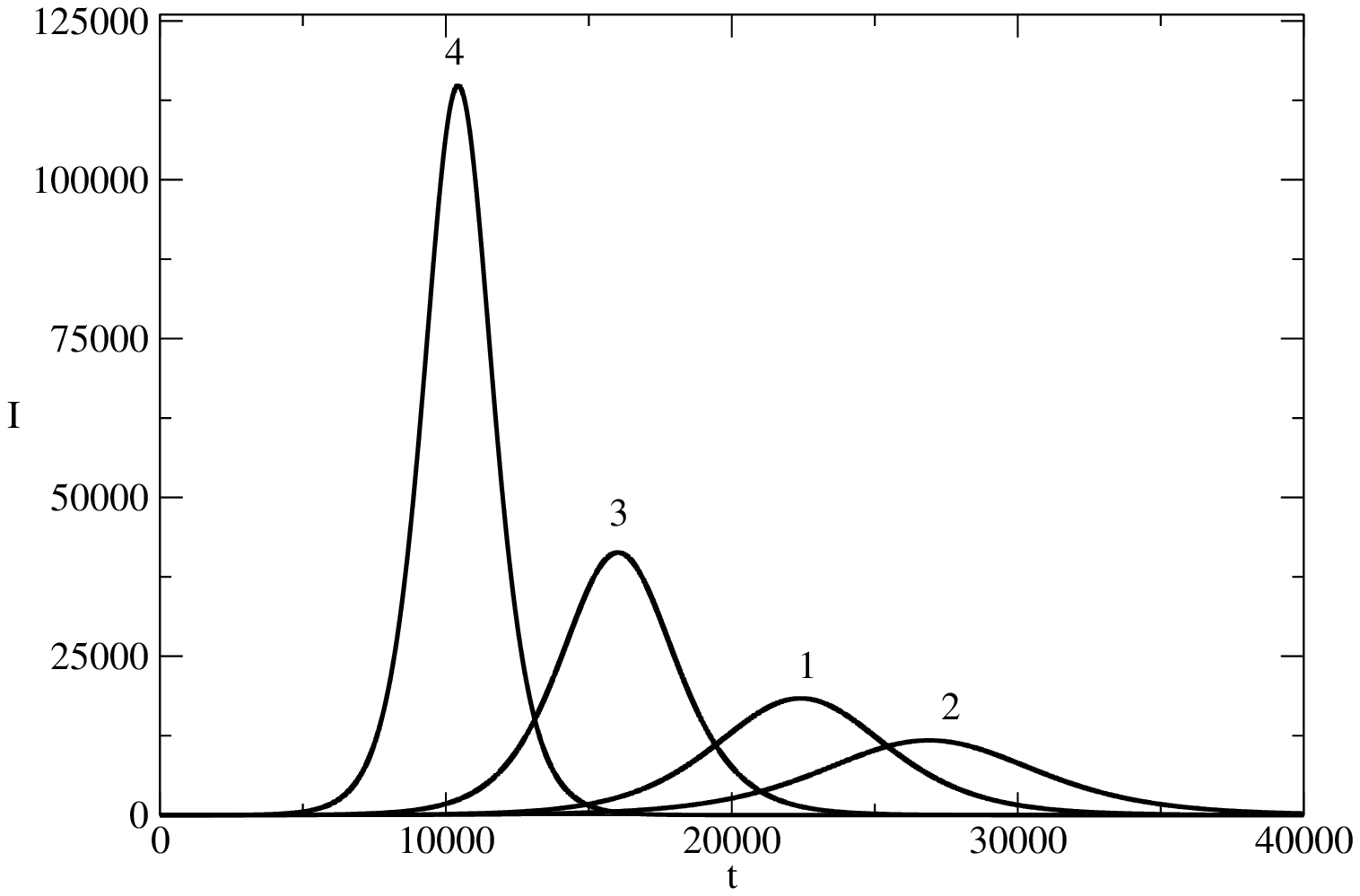}
\caption{Influence of the recovery rate $\rho$ on the  news wave. $I$ for the the basic solution  $N=10^7$, $S(0)=9 999 999$, $\tau=0.00825 $, $\rho=0.00775$ is
denoted by 1. In other curves we change only the values of $S(0)$ ($I(0)$). 
Curve 2: $\rho=0.00785$. Curve 3: $\rho=0.075$. Curve 4: $\rho = 0.7$.  }
\end{figure}
Figure 6 shows the influence of the changes of the value of the recovery rate $\rho$ on the news 
wave. The basic wave is marked by 1. We see that the increase in the recovery rate leads to smaller 
amplitude of the news wave. The decrease of the value of the recovery rate leads to an increase in 
the wave amplitude and a smaller time horizon $t_m$ of the wave.
\par 
\begin{figure}[htb!]
\centering
\includegraphics[angle=-90,scale=0.4]{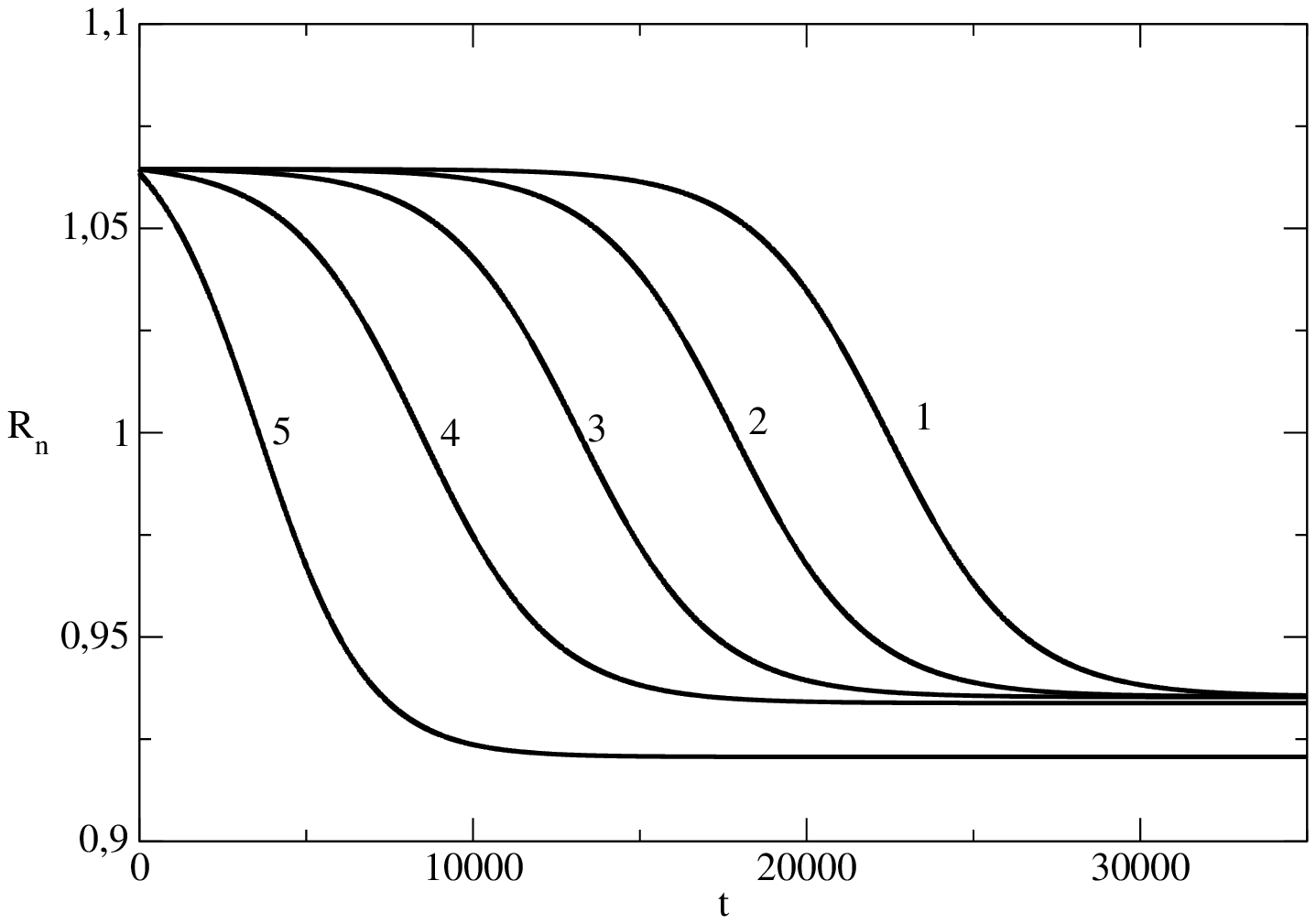}
\caption{Effective reproduction number $R_n$ for the news wave for
different values of $R_n$. $I$ for the basic solution  $N=10^7$, $S(0)=9 999 999$, $\tau=0.00825 $, $\rho=0.00775$ is
denoted by 1. In other curves we change only the values of $S(0)$ ($I(0)$). 
Curve 2: $I(0)=10$. Curve 3: $I(0)=100$. Curve 4: $I(0)=1000$. Curve 5: $I(0)=10 000$ }
\end{figure}
\par 
Figure 7 shows how the changes in the initial number $I(0)$
of individuals which spread the piece of news affects the effective reproduction number of the news wave. 
A smaller value of $I(0)$ leads to maintaining a larger $R_n$ for a longer time.  
\par 
\begin{figure}[htb!]
\centering
\includegraphics[angle=-90,scale=0.4]{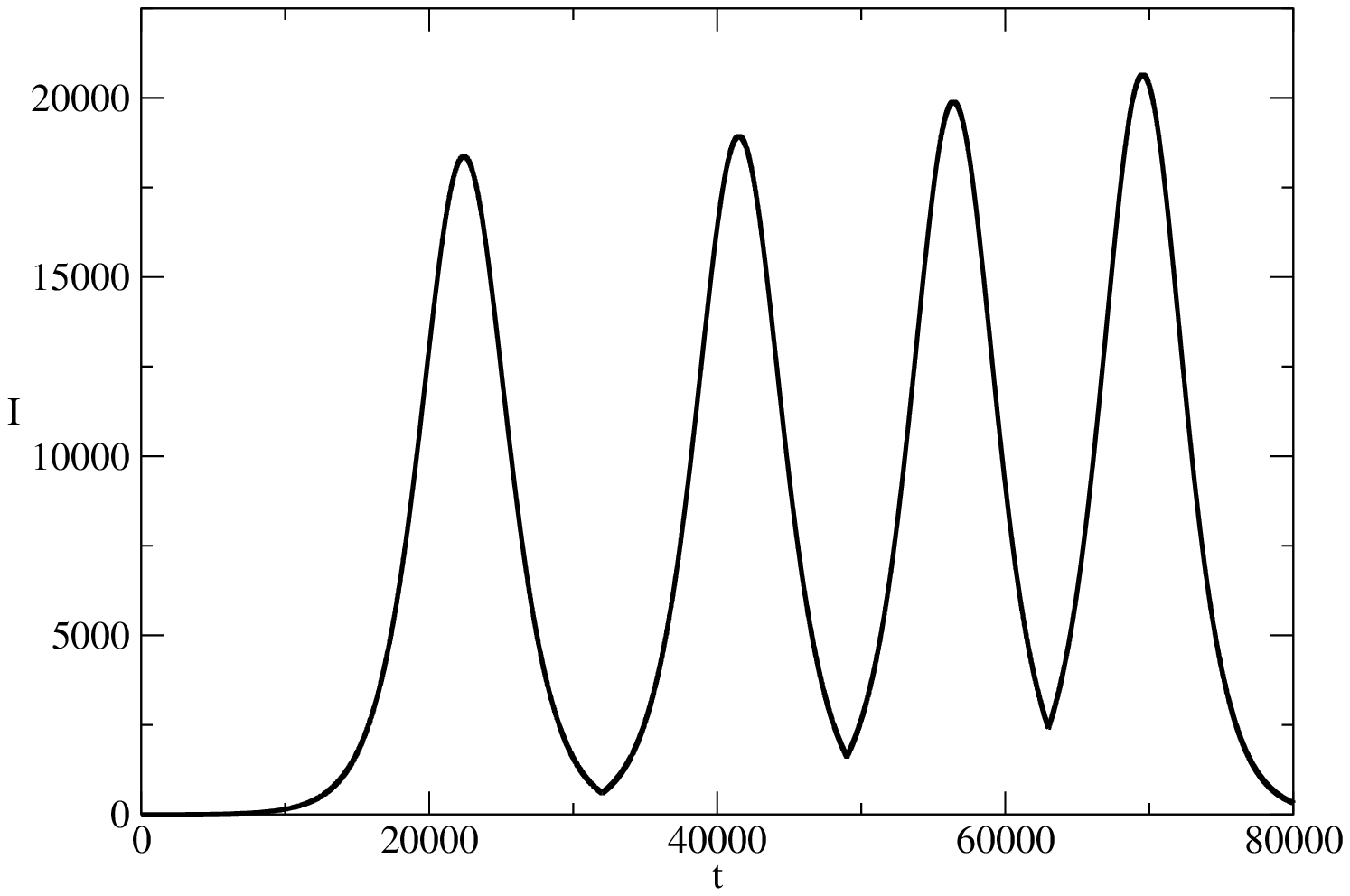}
\caption{A news wavetrain. The train is made by 4 waves, having the parameters of the basic solution 
from Figure 1. The wavetrain is an illustration of the possibility for long presence in  information 
space with increasing influence on the people in the same time.}
\end{figure}
Figure 8 illustrates the concept of news wavetrain. The news wavetrain is a possible technology for 
increasing influence by a sequence of news. Let us have a piece of news and this piece of news 
starts to spread in a population. $I(t)$ is the number of individuals who spread the news at the 
time $t$. At the time $t_1$ and before the vanishing of the first wave of the train, we have $I(t_1)$ 
spreaders of the corresponding news. One can use these active spreaders in order to start the second 
wave of the wavetrain at $t=t_1$. The second wave of the train can contain a piece of news which is 
similar but slightly different from the piece of news spread by the first wave. Figure 8 
illustrates the case when the initial number of spreaders of the second piece of news $I(t_1)$
is larger than the initial number $I(0)$ of spreaders of the first piece of news. This can lead to 
increase of the amplitude of the second wave of the news wavetrain with respect of the amplitude
of the first wave of the train.In the same manner, one can start the third wave of the train at 
$t=t_2$ and for $I(t_2)>I(t_1)$. This can lead to even larger amplitude of the third wave of the 
news wavetrain. The process can be continued (Fig. 8 shows a wavetrain containing four
news waves). In such a way, one can achieve two effects simultaneously.
\begin{enumerate}
\item 
One is present for a long time in the information part of the mind of 
the population and a possibility for  significant influence of this mind occurs.
\item
At the same time, one can affect a larger and larger part of this population.
\end{enumerate}
We will call the news wavetrain from Fig.8 an increasing news wavetrain because of the increasing 
amplitude of the news waves of the train. Construction of decreasing wavetrain is also possible. 
The third possibility is construction of a mixed amplitude news wavetrain where the amplitude of the 
waves of the train increases or decreases. Such wavetrain can have, for example, similar
shape as the wavetrains known from physics.
\par 
We note that an increasing news wavetrain can be constructed of news waves which have different values of the parameters $\rho$, $\theta$, and $I(t_n)$ at the beginning of the $n-1$-st wave of the train.
The increasing news wavetrain is a phenomenon which can be useful
in the case of the spreading of systems of ideas, in advertising, in  propaganda, etc.
\subsection{$D \ne 0$} 
In this case, we have to substitute $\alpha_0$, $\alpha_1$, $\alpha_2$ and $\theta$ from (\ref{sir8}) and (\ref{theta}) in (\ref{sirx12}) - (\ref{sirx14}). As the corresponding relationships become quite long,
we will not write them here. Instead of this we will evaluate the
correction in the case $D\ne 0$ in comparison to the case $D=0$. In order
to avoid the large calculations, we consider the specific case $t=-C$. In this case
from (\ref{sir19}) we obtain 
\begin{equation}\label{corr1}
R(-C) = - \frac{\alpha_1}{2 \alpha_2} + \frac{D}{E}
\end{equation}
$C$ is given by (\ref{sir11y}) and we have $C<0$ as the time $t$ has to be positive. $\theta$ is larger than $0$ and then we have that: 
$-\frac{1}{\theta}<\frac{\alpha_1 E - 2\alpha_2D}{2 \alpha_1 \alpha_2 D - \theta^2 E} <0$. Let us assume further that $D>0$, $E>0$, and $\alpha_1 >0$. $\alpha_2<0$ by definition. Thus, $2\alpha_1 \alpha_2 D - \theta^2E <0$ and $\alpha_1 E - 2 \alpha_2 D >0$. Then $C<0$ as required. From
$-\frac{1}{\theta}<\frac{\alpha_1 E - 2\alpha_2D}{2 \alpha_1 \alpha_2 D - \theta^2 E}$ we obtain $\frac{D}{E} <\frac{\theta}{-2 \alpha_2} $. The
upper bound of $D/E$ at $t=-C$ is $\frac{\theta}{-2 \alpha_2}$ and then
the upper bound of R(-C) is
\begin{equation}\label{corr2}
R(-C)\mid_{upper \ bound \ D \ne 0} = - \frac{\alpha_1}{2 \alpha_2} - \frac{\theta}{2\alpha_2}
\end{equation}
At the same value of $t=-C$ we have from (\ref{nw1})
\begin{eqnarray}\label{corr3}
R(-C)\mid_{D=0} = - \frac{\alpha_1}{2 \alpha_2} - \frac{\theta}{2 \alpha_2}
\tanh \left \{  \frac{\theta}{2} \Bigg [-\frac{2}{\theta} {\rm atanh} \Bigg[  \frac{\theta (\alpha_1 E - 2 \alpha_2 D)}{2 \alpha_1 \alpha_2 D - \theta^2 E} \Bigg]+\frac{2}{\theta} {\rm atanh} \Bigg(  -\frac{\alpha_1}{\theta} \right) \Bigg ] \Bigg \}
\end{eqnarray}
We note the following about  the changes in $R(t)$ for the case $D\ne 0$ 
in comparison to the case $D=0$. These changes accumulate and then vanish\
between $t=0$ and $t\to \infty$. The reason is that $R(t=0)=0$, and
$R(t\to \infty)\mid_{D=0} = R(t\to \infty)\mid_{D\ne 0} = -\frac{\alpha_1}{2 \alpha_2}$.
\subsection{The solution (\ref{sir24})}
The  quantities connected to this solution are given by (\ref{bern1}) - (\ref{bern5}).
The solution describes a specific kind of news wave. For this wave $R(0)=0$ and $R(t \to \infty) \approx N - S(0)/3$. In addition, $I(0) = N - S(0)$ as expected and $I(t \to \infty) \approx 0$.
This means that we have a news wave which decreases with increasing time and the amplitude of the
wave depends very much on the initial numbers of the spreaders of the piece of news. 
\par 
\begin{figure}[htb!]
\centering
\includegraphics[angle=-90,scale=0.4]{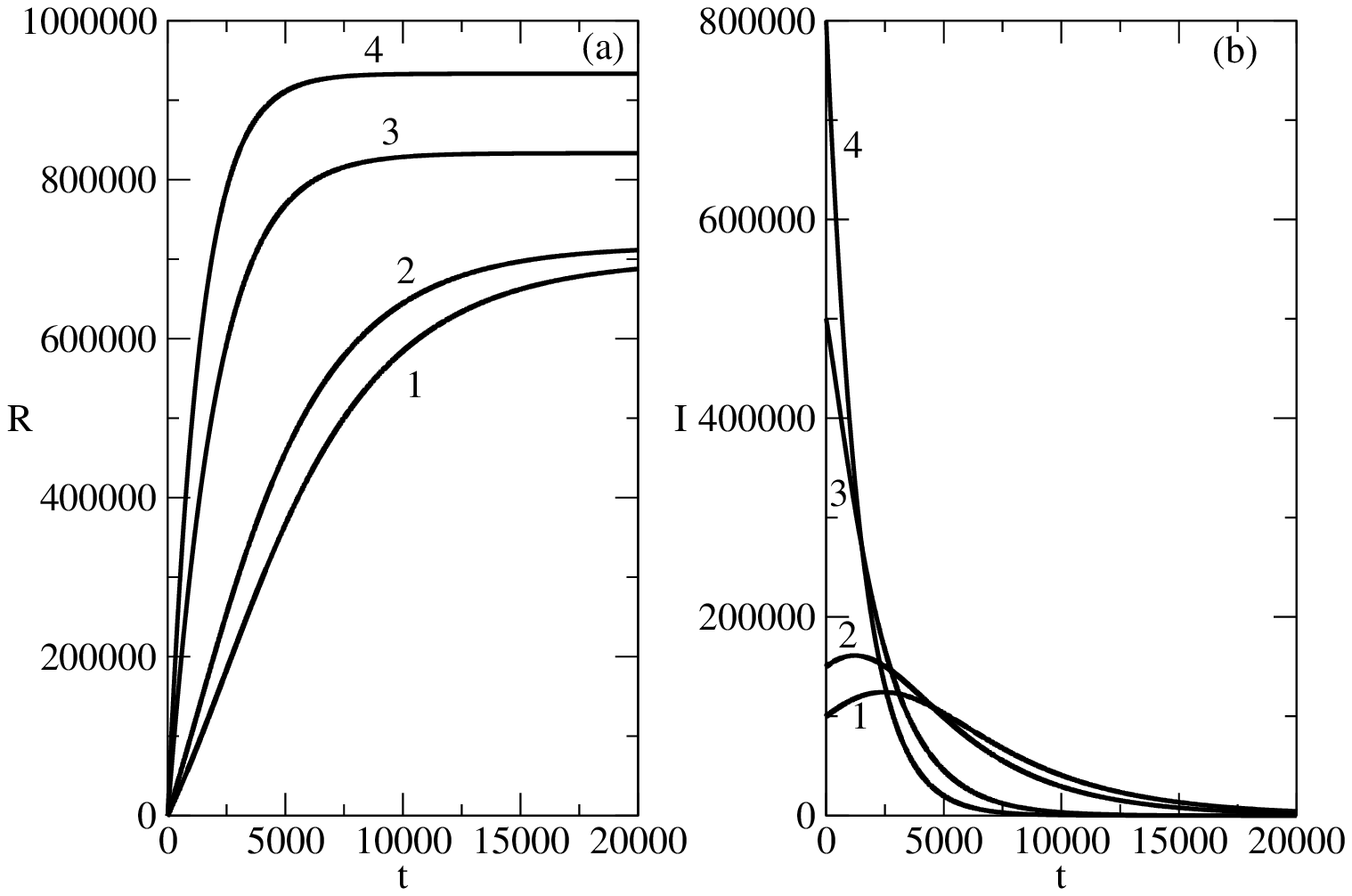}
\caption{The solution (\ref{sir24}). Figure (a) : Number of individuals affected by the news wave. Curve 1: Basic solution. $N= 1 000 000$, $S(0)= 900 000$, $\tau = 0.009$. Curves 2 - 4 show the influence of the change of the initial number of spreaders $I(0)=N-S(0)$ of the piece of news on
the number of individuals affected by the news wave. For the basic solution $I(0) = 10^5$ Curve 2:   $I(0)=1.5 \cdot 10^5$. Curve 3:$I(0)=5 \cdot 10^5$. Curve 4: $I(0)= 8 \cdot 10^5$. Figure (b): Influence of the number $I(0)$ of initial spreaders of the piece of news on the profile of the
news wave. Curve 1: Basic solution  $I(0) = 10^5$. Curve 2: $I(0) = 1.5 \cdot 10^5$, Curve 3: $I(0) = 5 \cdot 10^5$. Curve 4: $I(0) = 8 \cdot 10^5$ }
\end{figure}
\par 
Fig. 9 shows the influence of the initial number $I(0)$ of the individuals spreading the 
piece of news on the shape and the time horizon of the news wave of the kind (\ref{sir24}).
In order to have large wave of this kind, one has to start with a large number $I(0)$. For example,
in a population of $10^6$ individuals on has to start by $10^5$ initial spreaders of the
piece of news in order to reach about $7\cdot 10^5$ individuals (Curve 1 of Fig. 9(a)).
Fig. (9b) shows that the increase of the number of initial spreaders decreases the time horizon
of the wave and for large number of $I(0)$ the time horizon of the wave becomes $0$. Thus,
the waves of the kind (\ref{sir24}) require the presence of a large organization (in order to 
mobilize a large number of initial spreaders)and the waves are quite "dissipative" (the number of 
spreaders decrease very fast). 
\par 
\begin{figure}[htb!]
\centering
\includegraphics[angle=-90,scale=0.4]{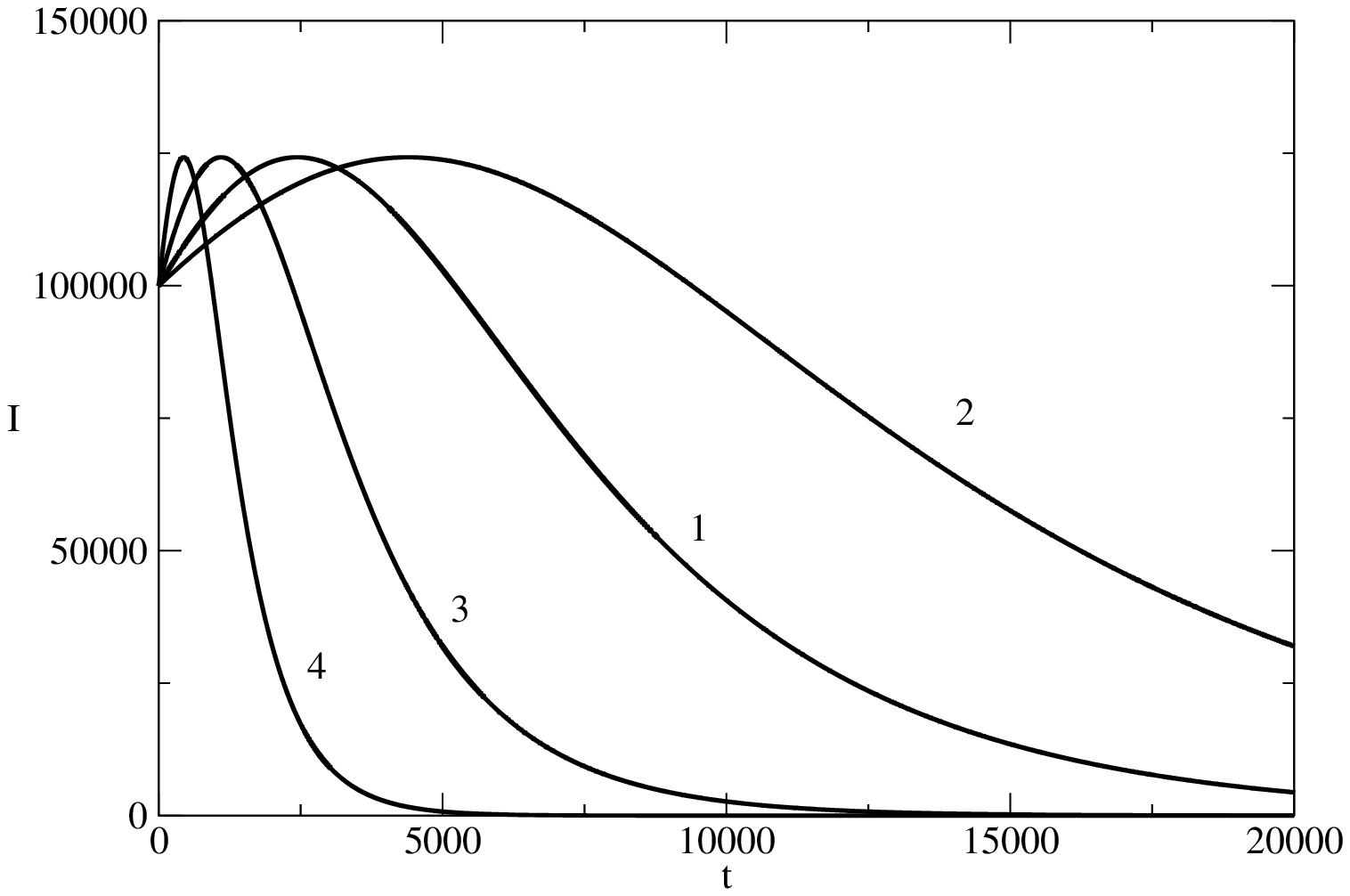}
\caption{The solution (\ref{sir24}). influence of the transmission rate $\tau$ on the number
of spreaders of the piece of news. Curve 1: Basic solution. $N= 1 000 000$, $S(0)= 900 000$, $\tau = 0.009$. Curve 2:   $\tau=0.0005$. Curve 3:$\tau = 0.002$. Curve 4: $\tau=0.005$.
}
\end{figure}
\par 
Fig. 10 shows the influence of the change of the transition rate $\tau$ on the shape and on 
the time horizon of the wave. The increase of the value of the transition rate leads to a decrease of
the time horizon of the wave and to the faster vanishing of the wave.  
\section{Discussion}
The discussed theory leads to several hints about the possibility for manipulation of the size, of the 
amplitude, and of time horizon of the news waves. They are as follows.
\begin{enumerate}
\item 
The organization of the process of initiation of a news wave is
important because the amplitude and the time horizon of the news wave
depend on the initial number of individuals which start to spread the
corresponding piece of news. If one wants to have a news wave which
possess a larger peak coming early in the time after the beginning
of the wave, then, one has to organize a larger number $I(0)$ of individuals which start to spread 
the piece of news. If one wants a
larger time horizon, then $I(0)$ must be smaller. But this will
lead to a news wave of smaller amplitude, i.e., the number of 
individuals affected by the news wave will be smaller.
\item 
In a region or country of a larger population the amplitude of the news wave will be larger and the time horizon will be longer in comparison to a region or country of a smaller population. 
Thus, the time of "life" of a piece of news in a large city or region
or country is expected to be longer than the time of "life" of the
same piece of news in a small town, small region or small country.
In addition, the piece of news will affect more individuals in a larger city, region or country than in a small town, region or country.
\item
The transmission rate $\tau$ strongly influences the amplitude and
the time horizon of the new wave. Thus, in order to achieve a news wave
of larger amplitude (more affected individuals by the piece of news) one
has to ensure a larger transmission rate (the corresponding population must be made more susceptible to the corresponding kind of news). But, the larger transmission rate also leads  to a shorter time horizon. In other words, the news wave of larger amplitude moves faster
through the population because of the higher permeability due to
the larger transmission rate.
\item
The increase of the recovery rate leads to a wave of smaller amplitude
and larger time horizon. Thus, if one wants to achieve a news wave of
larger amplitude, the recovery rate must be lowered. The appropriate
selection of the recovery rate can fix the position of the peak of
the news wave.
\item 
One can construct wave trains of news waves by use of pieces of
news with similar content. In such a case, one can use the number
of individuals who spread the piece of news at a given time as the population
of news spreaders who start to spread the next and slightly different piece of news. The wavetrains 
can be of three kinds. The news wavetrain
which could be of interest to advertising or propaganda is the 
increasing news wavetrain which allows to affect the population of
individuals whose number increases in time.
\end{enumerate}
\par 
\begin{figure}[htb!]
\centering
\includegraphics[angle=-90,scale=0.4]{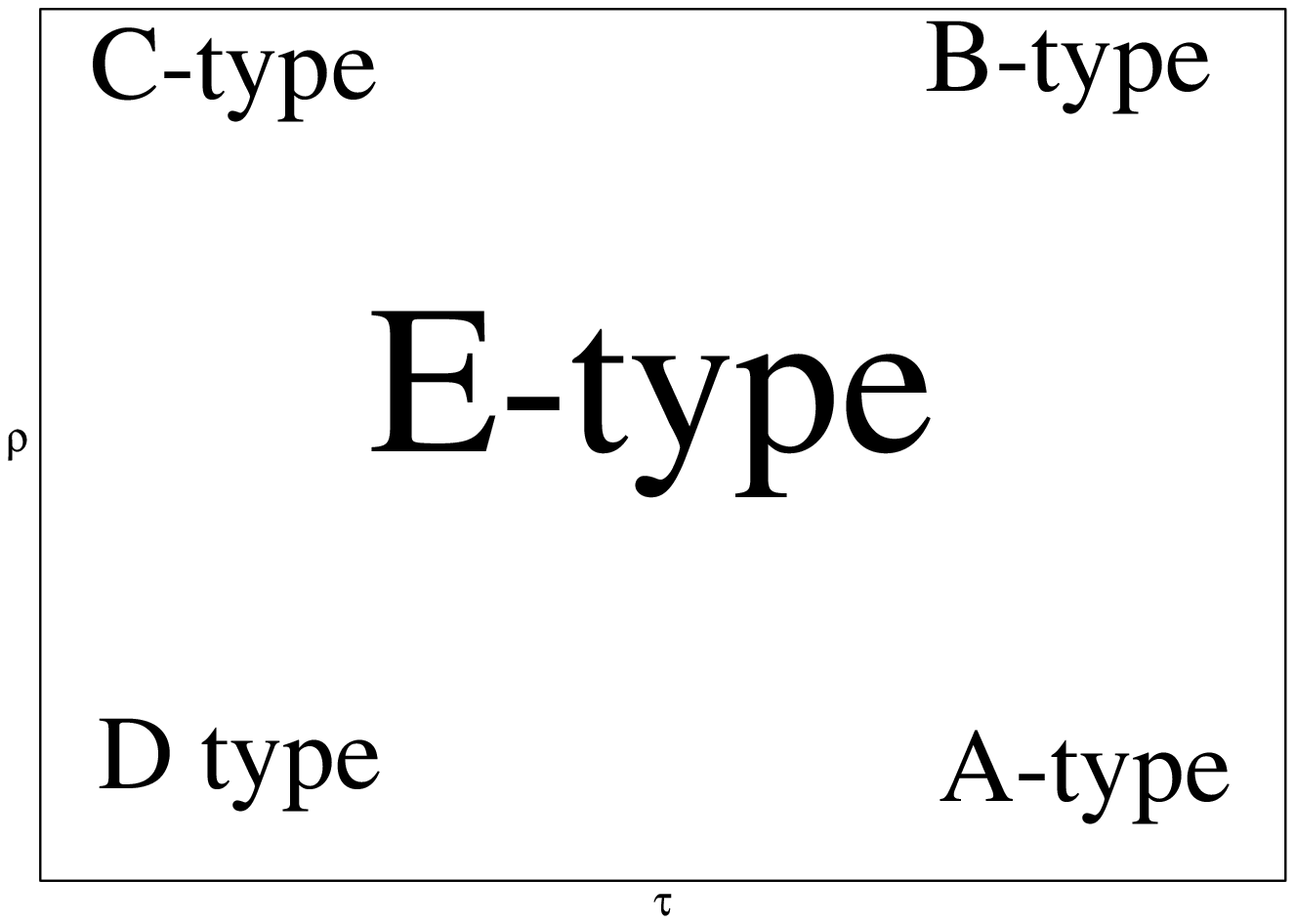}
\caption{Types of news waves  of the kind (\ref{nw1}) in the $\tau-\rho$-plane. A-type wave: large value of
$\tau$, small value of $\rho$. B-type wave: large value of $\tau$ large value of $\rho$. C-type wave:
small value of $\tau$, large value of $\rho$. D-type wave: small value of $\tau$, small value of
$\rho$. Type E-wave: values of $\tau$ and $\rho$ which can not be associated with one of the remaining
four types of waves. }
\end{figure}
\par 
We can consider 5 types of news waves of the kind (\ref{nw1}). Our classification will not be
based on characteristics like hardness of softness of the news or if the news is true or fake .
The classification will be quantitative and it will be based on the parameters $\tau$ and $\rho$.
We denote the types of news waves as: A-type, B-type, C-type, D-type, and E-type of news waves.
\par
The A-type of news waves has a large value of the transmission rate and a small value of the recovery rate.
This means that the population favours easy spread of the wave and the number of spreaders
of the corresponding piece of  news tends to decrease slowly. Such waves have a large amplitude and relatively 
small time horizon. The time horizon is small as the large amplitude compensates for the slow decrease of the
number of spreaders as $\frac{dR}{dt} = \rho I$. Thus, the A-type of news wave affects a large number of
individuals but lasts for a relatively short time period.
\par 
The B-type of news waves has a large value of the transmission rate and a large value for the recovery rate. 
The large value of $\tau$ favours a large amplitude and a small time horizon and the large value of $\rho$
 favours a smaller amplitude of the wave and a larger time horizon. Thus, the actual value of the
 wave amplitude and the time horizon of the B-type of news wave depend on the value of the ratio
 $\tau/\rho$.
 \par 
The C-type of news waves has a small value of the transmission rate and a large value of the
recovery rate. The small value of $\tau$ favours a  small amplitude and a large time horizon of the
news wave. The small amplitude and the large time horizon are also favoured by the large value
of $\rho$. Thus, the C-type of news wave will have a typically small amplitude and a large time horizon.
\par 
The D-type of news waves has a small value of the transmission rate and a  small value of the recovery rate.
The small transmission rate means that the population is not very permeable by the news wave. This
can be compensated by the small recovery rate, which means that the peace of the news affects the
individuals of the population for a longer time. Thus, we can construct a wave possessing  various values
of the amplitude and time horizon and this depends again on the ratio $\tau/\rho$ as in the case of the
B-type wave which however is connected to larger permeability and faster recovery in comparison to the
D-type of news wave.
\par 
Finally, the E-type of news wave is connected to the values of transmission rate and  recovery rate,
which are intermediate to the four other types of waves. By appropriate increase and decrease of the
values of $\tau$ and $\rho$, the E-type of news wave can be transformed to different type of wave.
For example, the increase of the values of $\tau$ and $\rho$ transforms the E-type of wave to
B-type of wave. The increase of $\tau$ and decrease of $\rho$ transforms the E-type of wave to
A-type of news wave. 
\par 
News waves are also possible on the basis on other solutions of equations from the studied
chain of equations. Such waves require a prescribed relationship for the ratio between the
transmission rate and the recovery rate. The increase of the transition rate for this wave
leads to a shorter time horizon and to a faster decrease of the number of spreaders of the
corresponding piece of news.  
\section{Concluding remarks}
In this article, we study the spread of news in a population. The basis of the study is the SIR model 
of the spread of epidemics. The model is reduced to a single equation which is associated with a chain of 
nonlinear differential equations which possess polynomial nonlinearities. By means of the Simple 
Equations Method (SEsM) we obtain exact solutions to several equations of this chain. We study the 
influence of the parameters of the model on the shape, peak, and the time horizon associated with the 
news waves. The presence of analytical solutions allows us to make many interesting conclusions about 
the possibility for change of the time horizon and maximum value of the wave. These conclusions give 
interesting hints for the practitioners.
\par 
The presented theory can be applied to various kinds of news, such as fake news, rumors, and for news in 
print media and news in social media. One can spread hard news or soft news and the corresponding
news wave can have parameters $\tau$ and $\rho$ which values belong to different regions. The 
difference in the values of the parameters of the news wave leads to different amplitude and
different time horizon of the wave. One can try to manipulate these parameters in order to obtain, for example, large amplitude short time horizon hard news wave or small amplitude long time horizon soft news wave. In addition, we consider the possibility for construction of wavetrains from several news waves. Three kinds of wavetrains are possible.
The most interesting for practice (in  areas such as advertising, propaganda, etc.) is the increasing wavetrain. This wavetrain allows
presence of a source of news for a long time in the minds of a
population and at the same time, the number of individuals influenced by the news, increases.
\begin{appendix}
\section{Several Remarks about the Simple Equations Method (SEsM)}\label{app1}
The Simple Equations Method (SEsM) was developed as a tool for use in
the modeling of numerous complex systems which range  from atomic 
chains and lattices, to  large biological systems  of
animals and groups of humans such as, for example, 
research groups and social networks or  economic systems.  
\cite{a1} - \cite{a4x}. As the complex systems are nonlinear
\cite{n1} - \cite{n8}, they must be studied by the
methods of  nonlinear time series analysis and by the
models used for such systems contain very often
nonlinear differential or difference equations \cite{t1} -
\cite{t10}. The analytical solutions of these nonlinear equations
are of large interest. Because of this, many researchers invest their 
efforts to obtain such solutions. We mention the Inverse Scattering T
ransform method \cite{ac} and the 
Method of Hirota \cite{hirota04}, as well as the Method of Simplest equation of Kudryashov
\cite{k05}, \cite{kl08}. The methodology of SEsM \cite{se1} - \cite{se7}  will be used in this text.
First steps in the development of SEsM have been made many years ago
 \cite{mv96}, \cite{mv5}. The work on the methodology continued in 
 \cite{1}, \cite{2},  and\cite{v10} and   the first version of the methodology called Modified  Method 
of Simplest Equation  (MMSE) was applied to problems  of population dynamics and  ecology \cite{vd10}. Further development of the methodology was made in \cite{vdk}, \cite{v11}  and numerous applications followed \cite{v11a}-\cite{vdv15}. Since 2018, we use
the current version of the methodology (SEsM). SEsM can use more than one simple equation \cite{vd18}.  The first 
discussion about SEsM was in  \cite{se1}. Further discussion on SEsM 
can be seen  in \cite{se2} - \cite{se4}, \cite{se7}. Applications 
of specific cases of SEsM can be seen in \cite{v21b} - \cite{v21c}. 
\par 
The general notation of SEsM is SEsM(n,m). This means that 
we have  to solve $n$  nonlinear differential equations by
means of the known solutions of $m$ simpler equations. The notation
for the case when we solve a single nonlinear differential equation is 
SEsM(1,m). The most used case up to now is SEsM(1,1): we  solve $1$ 
complicated nonlinear  differential equation by means of the known 
solutions of $1$ more  simple differential equation. SEsM(1,1) is also called    Modified 
Method of Simplest Equation.  There are many applications of  SEsM \cite{mmse1} - \cite{mmse5}.
\par 
The main idea of SEsM is to reduce the solved system 
of nonlinear differential equations to a specific form.
The solved system is
\begin{equation}\label{sesm1}
{\cal D}_i [u_{i1}(x,\dots,t), \dots, u_{in}(x,\dots,t) ] = 0, \ 
\ \ i=1,2,\dots,n.
\end{equation}
${\cal D}_i[u_{i1}(x,\dots,t), \dots, u_{in}(x,\dots,t), \dots]$ 
depends on the functions
$u_{i1}(x,\dots,t), \dots, u_{in}(x,\dots,t) $ and some of their 
derivatives.  The functions 
$u_{ik}$ can depend on several spatial coordinates.
The goal of SEsM is reduction of (\ref{sesm1}) to the specific form
\begin{equation}\label{sesm2}
\sum \limits_{i=1}^n  a_{ij}(\dots)E_{ij}=0, j=1,2,\dots, p_i. 
\end{equation}
Here, $E_{ij}$ are functions of the independent spatial  variables and of the time. $a_{ij}$ are 
relationships among the  parameters of the solved equations, parameters of the solutions 
and  the parameters of the solutions of the more simple equations.  
$p_i$ is a parameter which is characteristic  for the $i$-th 
solved equation. Note, that the relationships $a_{ij}$ 
contain only parameters whereas the spatial coordinates and the 
time are concentrated on the functions $E_{ij}$. If we manage to 
reduce the solved equations to the form (\ref{sesm1}) and then we 
set $a_{ij}=0$ we may obtain a nontrivial solution of the system 
of nonlinear differential equations. 
\par
In order to reduce (\ref{sesm1}) to (\ref{sesm2}) one can take 4 steps.
The first step is to use transformations which can remove the nonlinearities in (\ref{sesm1}) or can reduce these nonlinearities to
more treatable kind of nonlinearities (such as polynomial nonlinearities.
In the second step, one constructs the solutions of the solved equations as composite function of more simple equations. The kind of the
composite functions and the kind of the simple equations used are
determined in Step 3 in order to come to the relationships (\ref{sesm2}). Finally, one sets
\begin{equation}\label{sesm4}
a_{ij}(\dots) = 0
\end{equation}
 This leads to a system 
of nonlinear algebraic  relationships among the parameters of the 
solved equations,  the parameters of the composite  functions, 
and the parameters of the solutions of the more simple equations. 
Any nontrivial solution of (\ref{sesm4}) leads to a solution of 
the system (\ref{sesm1}).
\section{Several Exact Solutions of the Chain of Equations} \label{app2}
\par 
We apply SEsM below to the equation
\begin{equation}\label{sir10}
\frac{dR}{dt} = \sum \limits_{j=0}^M \alpha_j R^j,
\end{equation}
and we use as simple equations the differential equations
of Bernoulli and Riccati. We start by using  
the equation of Bernoulli
\begin{equation}\label{sir11}
\frac{dy}{dt} = py + q y^m, \ \ \ m=2,3,\dots,
\end{equation}
as a simple equation. By means of the transformation 
$y = u^{1/(1-m)}$ (\ref{sir11}) is reduced to 
a linear differential equation. This leads to the solution of the 
equation of Bernoulli as follows
\begin{equation}\label{sir12}
y(t) = \left \{ \frac{p}{  - q + C p  \exp[-(m-1)pt]} \right \}^{ 
\frac{1}{m-1}}.
\end{equation}
In (\ref{sir12}) $C$ is a constant of integration.
\par 
Then, we construct the composite function $R(y)$ as
\begin{equation}\label{sir13}
R (y) = \sum \limits_{l=0}^L \beta_l y^l,
\end{equation}
where $y(t)$ is the solution of the equation of Bernoulli 
(\ref{sir12}). At the following step of SEsM we obtain the balance 
equation. The 
presence of (\ref{sir11}) and (\ref{sir13}) fixes
the balance equation of (\ref{sir10}) to 
\begin{equation}\label{sir14}
m = 1+L(M-1).
\end{equation}
Thus, a specific solution of (\ref{sir10}) has the form
\begin{equation}\label{sir15}
R (t) = \sum \limits_{l=0}^L \beta_l \left \{ \frac{p}{  - q + C 
p  \exp \{-[L(M-1)]pt \}} \right \}^{ \frac{l}{L(M-1)}}.
\end{equation}
The parameters $\beta_l$, $p$, $q$ and $C$ are fixed by the 
solution of the system of nonlinear algebraic equations at the 
Step 4 of SEsM.
\par 
There is a specific case where we can obtain the general solution 
of (\ref{sir10}). This case is $M=2$. Here, (\ref{sir10}) becomes
\begin{equation}\label{sir16}
\frac{dR}{dt} = \alpha_2 R^2 + \alpha_1 R + \alpha_0.
\end{equation}
(\ref{sir16}) is an equation of Riccati kind. For this equation, 
we know the specific solution
\begin{equation}\label{sir17x}
R (t) =  - \frac{\alpha_1}{2 \alpha_2} - \frac{\theta}{
2 \alpha_2} \tanh \left[ \frac{\theta(t+C)}{2} \right], 
\end{equation}
where $\theta^2 = \alpha_1^2 - 4 \alpha_0 \alpha_2 >0$ and $C$ is 
a constant of integration. On the basis of the specific solution  
(\ref{sir17x}) of (\ref{sir16}),  we can write the
general solution of (\ref{sir16}) as $R = - \frac{\alpha_1}{2 
\alpha_2} - \frac{\theta}{2 \alpha_2} \tanh \left[ 
\frac{\theta(t+C)}{2} \right] + \frac{D}{v}$
where $D$ is a constant and  $v(t)$ is  the solution of the 
linear differential equation
\begin{equation}\label{sir17y}
\frac{dv}{dt} -  \theta \tanh \left[ \frac{\theta(t+C)}{2} 
\right] v = - \alpha_2 D
\end{equation}
The solution of (\ref{sir17y}) is
\begin{equation}\label{sir18}
v = \cosh^2\left[ \frac{\theta(t+C}{2} \right] \left \{ E - 
\frac{2 \alpha_2 D}{\theta}
\tanh \left[ \frac{\theta(t+C}{2} \right] \right \},
\end{equation}
where $E$ is a constant of integration. Then, the general 
solution of the equation (\ref{sir16}) is (\ref{sir19}).
\par 
Let us now obtain the form of several solutions of the kind 
(\ref{sir15}). For $M=2$ we have the general solution 
(\ref{sir19}) of the corresponding equation (\ref{sir16}). Thus, 
we start from $M=3$. We note, that the specific case for this text is
that we consider only solutions for
which $L=1$. For these solutions, it follows from (\ref{sir15})
that
\begin{eqnarray}\label{gsol}
R (t) = \beta_0 + \beta_1 \left \{ \frac{p}{  - q + C 
p  \exp \{-[(M-1)]pt \}} \right \}^{ \frac{1}{(M-1)}}.
\end{eqnarray}
For $M=3$, we solve the equation
The equation we have to solve is
\begin{equation}\label{sir20}
\frac{dR}{dx} = \alpha_3 R^3 + \alpha_2 R^2 + \alpha_1 R + 
\alpha_0.
\end{equation}
The balance equation is
$m=3$. This fixes the form of the simple equation of Bernoulli 
for this case:
$\frac{dy}{dx} = py + q y^{3}$. The substitution of the 
last relationships in (\ref{sir20})
leads to the following system of nonlinear algebraic 
relationships
\begin{eqnarray}\label{sir21}
  q - \alpha_3 \beta_1^2 &=&0, \nonumber \\
 3 \alpha_3 \beta_0 + \alpha_2 &=&0, \nonumber \\
p-3 \alpha_3 \beta_0^2 - \alpha_1 - 2 \alpha_2 \beta_0 &=& 0, \nonumber \\
-\alpha_1 \beta_0 - \alpha_3 \beta_0^3 - \alpha_0 - \alpha_2 \beta_0^2 &=& 0 . 
\end{eqnarray}
The solution of (\ref{sir21}) is
\begin{eqnarray}\label{sir22}
p= \frac{3 \alpha_1 \alpha_3 - \alpha_2^2}{3 \alpha_3}; \ \ \ q = \alpha_3 \beta_1^2;  \ \ \ 
\beta_0 = - \frac{\alpha_2}{3 \alpha_3}; \ \ \
\alpha_0 = \frac{\alpha_2(9 \alpha_1 \alpha_3 - 2 \alpha_2^2)}{27 \alpha_3^2}.
\end{eqnarray}
Thus, the equation
\begin{equation}\label{sir23}
\frac{dR}{dx} = \alpha_3 R^3 + \alpha_2 R^2 + \alpha_1  R +\frac{\alpha_2(9 \alpha_1 \alpha_3 - 2 \alpha_2^2)}{27 \alpha_3^2},
\end{equation}
has the specific exact analytical solution (\ref{sir24}). Note, that we have one relationship
among the parameters $a_0$, $a_1$,$a_2$, $a_3$ in (\ref{sir22}).
\par
Next, we consider the case $M=4$. We have to solve the equation
\begin{equation}\label{sir29}
\frac{dR}{dx} = \alpha_4 R^4 + \alpha_3 R^3 + \alpha_2 R^2 + 
\alpha_1 R + \alpha_0.
\end{equation}
 The solution is of the kind (\ref{gsol}) and from (\ref{sir14}) 
 we have the balance equation
$m=4$. This fixes the form of the simple equation of Bernoulli 
for this case:
$\frac{dy}{dx} = py + q y^{4}$. The substitution of the 
above relationships in (\ref{sir29})
leads to the following system of nonlinear algebraic 
relationships
\begin{eqnarray}\label{sir30}
q-\alpha_4 \beta_1^3 &=& 0, \nonumber \\
\alpha_3 + 4 \alpha_4 \beta_0&=&0, \nonumber \\
3 \alpha_3 \beta_0 + \alpha_2  + 6 \alpha_4 \beta_0^2 &=& 0, \nonumber \\ 
p - 3 \alpha_3 \beta_0^2 - 2 \alpha_2 \beta_0 - 4 \alpha_4 \beta_0^3 - \alpha_1 &=& 0, \nonumber \\
-\alpha_1 \beta_0 -\alpha_4 \beta_0^4 - \alpha_0 - \alpha_3 \beta_0^3 - \alpha_2 \beta_0^2 &=& 0. \nonumber \\
\end{eqnarray}
The solution of (\ref{sir30}) is
\begin{eqnarray}\label{sir31}
\beta_0 = - \frac{\alpha_3}{4 \alpha_4}; \ \ \ q = \alpha_4 \beta_1^3; \ \ \ p = - \frac{\alpha_3^4 - 256 \alpha_0 \alpha_4^3}{64 \alpha_3 \alpha_4^2}; \ \ \  
\alpha_1 = \frac{3 \alpha_3^4 + 256 \alpha_0 \alpha_4^3}{64 \alpha_4^2 \alpha_3}; 
\ \ \  \alpha_2 = 
\frac{3\alpha_3^2}{8 \alpha_4}
\end{eqnarray}
Thus, the equation
\begin{equation}\label{sir32}
\frac{dR}{dx} = \alpha_4 R^4 + \alpha_3 R^3 +  
\frac{\alpha_3^2}{8 \alpha_4} R^2 + \frac{3 \alpha_3^4 + 256 \alpha_0 \alpha_4^3}{64 \alpha_4^2 \alpha_3} R + \alpha_0,
\end{equation}
has the solution (\ref{sir33}). Note, that we have two relationships
among the parameters $a_0$, $a_1$,$a_2$, $a_3$, $a_4$ in (\ref{sir31}).
\par 
Next, we consider the case $M=5$. We have to solve the equation
\begin{equation}\label{sir34}
\frac{dR}{dx} = \alpha_5 R^5 + \alpha_4 R^4 + \alpha_3 R^3 + 
\alpha_2 R^2 + \alpha_1 R + \alpha_0.
\end{equation}
 The solution is of the kind (\ref{sir13}) and from \ref{sir14}) 
 we have the balance equation
$m=5$. This fixes the form of the simple equation of Bernoulli 
for this case:
$\frac{dy}{dx} = py + q y^{5}$. The substitution of the 
last relationships in (\ref{sir34})
leads to the following system of nonlinear algebraic 
relationships
\begin{eqnarray}\label{sir35}
q- \alpha_5 \beta_1^4 &=& 0, \nonumber \\
\alpha_4 + 5 \alpha_5 \beta_0 &=& 0, \nonumber \\
\alpha_3 + 4 \alpha_4 \beta_0 + 10 \alpha_5 \beta_0^2&=& 0, \nonumber \\
\alpha_2 + 3 \alpha_3 \beta_0 + 6 \alpha_4 \beta_0^2 -10 \alpha_5 \beta_0^3 &=& 0, \nonumber \\
3 \alpha_3 \beta_0^2 + 4 \alpha_4 \beta_0^3 + 5 \alpha_5 \beta_0^4 + 2 \alpha_2 \beta_0 -p + \alpha_1 &=& 0, \nonumber \\
\alpha_3 \beta_0^3 + \alpha_0 + \alpha_5 \beta_0^5 - \alpha_1 \beta_0 - \alpha_2 \beta_0^2 - \alpha_4 \beta_0^4 &=& 0 .
\end{eqnarray}
The solution of (\ref{sir35}) is
\begin{eqnarray}\label{sir36}
q =\alpha_5 \beta_1^4; \ \ \ \
p = \frac{-\alpha_4^5+3125 \alpha_0 \alpha_5^4}{625 \alpha_4 \alpha_5^3}; \ \ \
\beta_0 = - 
\frac{\alpha_4}{5 \alpha_5}; \ \ \ \alpha_1 = \frac{4\alpha_4^5 + 3125 \alpha_0 \alpha_5^4}{625 \alpha_5^3 \alpha_4}; \nonumber \\ 
\alpha_2 = \frac{2 \alpha_4^3}{25 \alpha_5^2}; \ \ \ \alpha_3= 
\frac{2 \alpha_4^2}{5 \alpha_5}
\end{eqnarray}
Thus, the equation
\begin{equation}\label{sir37}
\frac{dR}{dx} = \alpha_5 R^5 + \alpha_4 R^4+\frac{2 \alpha_4^2}{5 
\alpha_5} R^3 + \frac{2 \alpha_4^3}{25 \alpha_5^2} R^2 + 
\frac{4\alpha_4^5 + 3125 \alpha_0 \alpha_5^4}{625 \alpha_5^3 \alpha_4} R + \frac{\alpha_4^5}{3125 
\alpha_5^4},
\end{equation}
has a specific solution (\ref{sir38}.)  Note, that we have two relationships
among the parameters $a_0$, $a_1$,$a_2$, $a_3$, $a_4$, $a_5$ in (\ref{sir36}).
\par 
The obtaining of exact solutions of the chain of equations can be 
continued. Below we focus on the epidemic waves connected to the 
SIR model. The additional exact solutions of the chain of the 
equations will be discussed elsewhere.
\end{appendix}

\end{document}